\documentclass[11pt]{article}
\usepackage[utf8]{inputenc}
\usepackage{amsfonts}
\usepackage{amsmath}
\usepackage{amssymb}
\usepackage{indentfirst}
\usepackage{graphicx}
\usepackage{hyperref}
\usepackage{cite}
\usepackage{color}
\usepackage{xcolor}

\usepackage{microtype}
\usepackage[normalem]{ulem}

\numberwithin{equation}{section}
\allowdisplaybreaks

\begin{document}

\baselineskip=18pt 
\baselineskip 0.6cm
\begin{titlepage}
\vskip 4cm

\begin{center}
\textbf{\LARGE{Non-relativistic three-dimensional supergravity theories and semigroup expansion method}}
\par\end{center}{\LARGE \par}

\begin{center}
	\vspace{1cm}
	\textbf{Patrick Concha}$^{\ast}$,
	\textbf{Marcelo Ipinza}$^{\S}$,
    \textbf{Lucrezia Ravera}$^{\star , \ddag}$,
	\textbf{Evelyn Rodríguez}$^{\dag}$,
	\small
	\\[6mm]
	$^{\ast}$\textit{Departamento de Matemática y Física Aplicadas, }\\
	\textit{ Universidad Católica de la Santísima Concepción, }\\
\textit{ Alonso de Ribera 2850, Concepción, Chile.}
	\\[3mm]
	$^{\S}$\textit{Instituto
	de Física, Pontificia Universidad Católica de Valparaíso, }\\
	\textit{ Casilla 4059, Valparaiso-Chile.}
	\\[2mm]
    $^{\star}$\textit{DISAT, Politecnico di Torino, }\\
	\textit{ Corso Duca degli Abruzzi 24, 10129 Torino, Italy.}
	\\[3mm]
	$^{\ddag}$\textit{INFN, Sezione di Torino, }\\
	\textit{ Via P. Giuria 1, 10125 Torino, Italy.}
	\\[3mm]
	$^{\dag}$\textit{Departamento de Ciencias, Facultad de Artes Liberales,} \\
	\textit{Universidad Adolfo Ibáñez, Viña del Mar-Chile.} \\[5mm]
	\footnotesize
	\texttt{patrick.concha@ucsc.cl},
	\texttt{marcelo.calderon@pucv.cl},
    \texttt{lucrezia.ravera@polito.it},
	\texttt{evelyn.rodriguez@edu.uai.cl}
	\par\end{center}
\vskip 20pt
\centerline{{\bf Abstract}}
\medskip
\noindent In this work we present an alternative method to construct diverse  non-relativistic Chern-Simons supergravity theories in three spacetime dimensions. To this end, we apply the Lie algebra expansion method based on semigroups to a supersymmetric extension of the Nappi-Witten algebra. Two different families of non-relativistic superalgebras are obtained, corresponding to generalizations of the extended Bargmann superalgebra and extended Newton-Hooke superalgebra, respectively. The expansion method considered here allows to obtain known and new non-relativistic supergravity models in a systematic way. In particular, it immediately provides an invariant tensor for the expanded superalgebra, which is essential to construct the corresponding Chern-Simons supergravity action. We show that the extended Bargmann supergravity and its Maxwellian generalization appear as particular subcases of a generalized extended Bargmann supergravity theory. In addition, we demonstrate that the generalized extended Bargmann and generalized extended Newton-Hooke supergravity families are related through a contraction process. 

\end{titlepage}\newpage {\baselineskip=12pt \tableofcontents{}}

\section{Introduction}\label{intro}

The formulation of a non-relativistic (NR) three-dimensional supergravity theory has recently been approached in \cite{Andringa:2013mma} and subsequently developed in \cite{Bergshoeff:2015ija,Bergshoeff:2016lwr}. These last two years, the construction of NR supergravity actions has received a growing interest \cite{Ozdemir:2019orp,deAzcarraga:2019mdn,Ozdemir:2019tby,Concha:2019mxx,Concha:2020tqx} considering different procedures. Such supergravity models correspond to supersymmetric extensions of diverse NR gravity theories. Unlike their bosonic counterparts, the construction of NR supergravity theories remains as a challenging task mainly motivated by the diverse applications of these models in the context of holography and relativistic field theory. 

The first proposed NR supergravity theory corresponds to a supersymmetric extension of the Newton-Cartan gravity which was obtained by gauging a Bargmann superalgebra \cite{Andringa:2013mma}. At the bosonic level, the Newton-Cartan formalism allows to formulate in a geometric way a Newtonian gravity model that resembles General Relativity \cite{Cartan1,Cartan2}. Newton-Cartan gravity theories have been largely studied and extended with diverse purposes \cite{Duval:1983pb,Duval:1984cj,Duval:2009vt,Andringa:2010it,Banerjee:2014nja,Banerjee:2016laq,Bergshoeff:2017dqq,Aviles:2018jzw,Aviles:2019xed,Chernyavsky:2019hyp,Concha:2019lhn,Harmark:2019upf,Hansen:2020pqs,Ergen:2020yop,Kasikci:2020qsj,Concha:2020sjt}. Remarkably, Newton-Cartan geometry has been useful to approach strongly coupled condensed matter systems \cite{Son:2008ye,Balasubramanian:2008dm,Kachru:2008yh,Bagchi:2009my,Bagchi:2009pe,Christensen:2013lma,Christensen:2013rfa,Hartong:2014oma,Hartong:2014pma,Hartong:2015wxa,Taylor:2015glc} and NR effective field theories \cite{Hoyos:2011ez,Son:2013rqa,Abanov:2014ula,Geracie:2015dea,Gromov:2015fda}. Nevertheless, an action principle for a NR supergravity theory requires to consider an approach different from the Newton-Cartan supergravity one. To this end, a Chern-Simons (CS) formalism was considered in \cite{Bergshoeff:2016lwr} to construct a three-dimensional NR supergravity action invariant under an extended Bargmann superalgebra. Such superalgebra can be seen as the supersymmetric extension of the extended Bargmann algebra \cite{Grigore:1993fz,Bose:1994sj,Duval:2000xr,Jackiw:2000tz,Papageorgiou:2009zc}. The extended Bargmann superalgebra admits an invariant bilinear form which ensures the proper construction of a well-defined CS action. Furthermore, the extended Bargmann gravity differs from the Newton-Cartan gravity at the matter coupling level, allowing all components of the Ricci tensor to be non-vanishing. On the other hand, the CS formalism has the advantage of offering a gauge-invariant action, this being an interesting three-dimensional toy model \cite{Achucarro:1987vz,Witten:1988hc,Zanelli:2005sa}.

To go beyond Poincaré supergravity is a natural step to explore more general supergravity theories. Analogously, at the NR level, it is possible to extend the extended Bargmann supergravity to approach other features. The inclusion of a cosmological constant in a NR supergravity theory was presented in \cite{Ozdemir:2019tby} which, in the flat limit, reproduces the extended Bargmann supergravity. The NR action is based on the extended Newton-Hooke superalgebra which appears as a Lie algebra expansion \cite{deAzcarraga:2002xi} of the $\mathcal{N}=2$ AdS superalgebra. More recently, a Maxwellian generalization of the extended Bargmann supergravity and its extension to an enlarged extended Bargmann supergravity were studied in \cite{Concha:2019mxx} and \cite{Concha:2020tqx}, respectively. They can be seen as supersymmetric extensions of the Maxwellian extended Bargmann (MEB) \cite{Aviles:2018jzw} and enlarged extended Bargmann (EEB) gravity theories \cite{Concha:2019lhn}. While the MEB supergravity action has been obtained by hand, the EEB supergravity theory has been found from the relativistic $\mathcal{N}=2$ AdS-Lorentz superalgebra through the semigroup expansion method \cite{Izaurieta:2006zz}. 

The Lie algebra expansion procedure has been introduced in \cite{Hatsuda:2001pp} and subsequently developed by expanding Maurer-Cartan forms \cite{deAzcarraga:2002xi,deAzcarraga:2007et}. An expansion method based on semigroups ($S$-expansion) has been then introduced in \cite{Izaurieta:2006zz} and subsequently studied in \cite{Caroca:2011qs,Andrianopoli:2013ooa,Artebani:2016gwh,Ipinza:2016bfc,Inostroza:2017ezc,Inostroza:2018gzd}. Within the $S$-expansion procedure the expanded (super)algebra is obtained by combining the structure constants of a Lie (super)algebra with the multiplication law of a semigroup $S$. In addition, the $S$-expansion method provides us with the non-vanishing components of the invariant tensor for the expanded (super)algebra, which are crucial to construct CS actions. The $S$-expansion mechanism not only has been useful at the NR level\footnote{Application at the NR level of the Lie algebra expansion method considering the Maurer-Cartan equations can be found in \cite{Bergshoeff:2019ctr,deAzcarraga:2019mdn,Romano:2019ulw,Kasikci:2020qsj,Fontanella:2020eje}.} \cite{Concha:2019lhn,Penafiel:2019czp,Gomis:2019nih,Bergshoeff:2020fiz,Concha:2020sjt,Concha:2020ebl} but also to obtain novel relativistic symmetries \cite{Izaurieta:2009hz,Diaz:2012zza,Concha:2013uhq,Salgado:2014qqa,Caroca:2017izc}, superalgebras \cite{Izaurieta:2006aj,Fierro:2014lka,Concha:2014tca,Concha:2016zdb,Banaudi:2018zmh,Concha:2018jxx}, and asymptotic symmetries \cite{Caroca:2017onr,Caroca:2018obf,Caroca:2019dds}, among others.

In this work, we present an alternative procedure to construct various NR supergravity theories by considering the $S$-expansion method. We extend the results obtained in \cite{Concha:2019lhn,Penafiel:2019czp} in which diverse NR symmetries are obtained by expanding the Nappi-Witten algebra. The Nappi-Witten symmetry was introduced in \cite{Nappi:1993ie,Figueroa-OFarrill:1999cmq} and can be seen as a central extension of the homogeneous part of the Galilei algebra. Here, we apply the $S$-expansion procedure to the Nappi-Witten superalgebra introduced in \cite{Concha:2020tqx} to obtain known and new NR superalgebras.
We get two families of NR superalgebras by considering two different semigroup families. In particular, we first show that the extended Bargmann superalgebra and its generalizations can be obtained as an $S$-expansion of the super Nappi-Witten algebra. Then, using the same method but different semigroups, we get the extended Newton-Hooke superalgebra and its generalizations. The construction of NR CS supergravity actions for each NR superalgebra is also presented. Interestingly, we prove that the extended Bargmann supergravity along with its Maxwellian version correspond to particular subcases of a generalized extended Bargmann supergravity theory. Furthermore, they can alternatively be obtained as an Inönü-Wigner (IW) contraction \cite{Inonu:1953sp} of the generalized extended Newton-Hooke supergravity theory presented here.  Our construction offers a systematic way to obtain different NR supergravity models which are supersymmetric extensions of distinct NR gravity theories.

The organization of the paper is as follows: In Section \ref{snw}, we discuss the supersymmetric extension of the Nappi-Witten algebra. In Section \ref{geb}, we obtain the extended Bargmann supergravity theory, its Maxwellian version, and further generalizations by applying the $S$-expansion method to the super Nappi-Witten algebra and its invariant tensor. In Section \ref{gnh}, we recover an extended Newton-Hooke supergravity along with generalizations of the latter by considering different semigroups. Section \ref{concl} is devoted to some concluding remarks.

\section{Nappi-Witten superalgebra and Chern-Simons action}\label{snw}

A supersymmetric extension of the Nappi-Witten algebra was recently introduced in \cite{Concha:2020tqx}. In addition to the Nappi-Witten generators $\lbrace J,G_a,S\rbrace$, it contains two additional bosonic generators $\lbrace T_1,T_2\rbrace$ and three Majorana fermionic generators given by $Q_{\alpha}^{+}$, $Q_{\alpha}^{-}$, and $R_{\alpha}$ (with $\alpha,\beta,\ldots=1,2$). The extra bosonic content assures not only the Jacobi identities of the superalgebra but also the non-degeneracy of the invariant tensor. The super Nappi-Witten generators satisfy the following (anti-)commutation relations:\footnote{In this work, we are using a mostly plus metric.}
\begin{eqnarray}
\left[ J,G_{a}\right] &=&\epsilon _{ab}G_{b}\,,\qquad \qquad \ \ \ \quad \left[
G_{a},G_{b}\right] =-\epsilon _{ab}S\,,  \notag \\
\left[ J,Q_{\alpha }^{\pm }\right] &=&-\frac{1}{2}\left( \gamma _{0}\right)
_{\alpha }^{\,\ \beta }Q_{\beta }^{\pm }\,,\quad \ \ \ \ \ \ \left[
J,R_{\alpha }\right] =-\frac{1}{2}\left( \gamma _{0}\right) _{\alpha }^{%
\text{ }\beta }R_{\beta }\,,  \notag \\
\left[ G_{a},Q_{\alpha }^{+}\right] &=&-\frac{1}{2}\left( \gamma _{a}\right)
_{\alpha }^{\,\ \beta }Q_{\beta }^{-}\,,\quad \ \ \ \left[ G_{a},Q_{\alpha
}^{-}\right] =-\frac{1}{2}\left( \gamma _{a}\right) _{\alpha }^{\text{ }%
\beta }R_{\beta }\,,  \notag \\
\left[ S,Q_{\alpha }^{+}\right] &=&-\frac{1}{2}\left( \gamma _{0}\right)
_{\alpha }^{\text{ }\beta }R_{\beta }\,,\quad \ \ \quad \left[
T_{1},Q_{\alpha }^{\pm}\right] = \pm \frac{1}{2}\left( \gamma _{0}\right) _{\alpha
\beta }Q_{\beta }^{\pm}\,,  \notag \\
\left[ T_{2},Q_{\alpha }^{+}%
\right] &=& \frac{1}{2}\left( \gamma _{0}\right) _{\alpha \beta }R_{\beta } \,,\qquad \quad \,\,\,\, \left[ T_{1},R_{\alpha }\right] = \frac{1}{2}\left( \gamma _{0}\right)
_{\alpha \beta }R_{\beta } \,, \notag \\
\left\{ Q_{\alpha }^{+},Q_{\beta }^{-}\right\} &=&-\left( \gamma
^{a}C\right) _{\alpha \beta }G_{a}\,,  \notag \\
\left\{ Q_{\alpha }^{+},Q_{\beta }^{+}\right\} &=&-\left( \gamma
^{0}C\right) _{\alpha \beta }J-\left( \gamma ^{0}C\right) _{\alpha \beta
}T_{1}\,,\text{ \qquad }  \notag \\
\left\{ Q_{\alpha }^{-},Q_{\beta }^{-}\right\} &=&-\left( \gamma
^{0}C\right) _{\alpha \beta }S {+} \left( \gamma ^{0}C\right) _{\alpha \beta
}T_{2}\,,  \notag \\
\left\{ Q_{\alpha }^{+},R_{\beta }\right\} &=&-\left( \gamma ^{0}C\right)
_{\alpha \beta }S-\left( \gamma ^{0}C\right) _{\alpha \beta }T_{2}\,,  \label{sNW}
\end{eqnarray}%
where $a,b,\ldots=1,2$, $\epsilon_{ab}=\epsilon_{0ab}$, $\epsilon^{ab}=\epsilon^{0ab}$, $\gamma^0$ and $\gamma^a$ are the Dirac gamma matrices in three dimensions, and $C$ is the charge conjugation matrix. Let us note that the Nappi-Witten superalgebra (\ref{sNW}) can be decomposed
in a bosonic subspace $V_{0}=\left\{ J,G_{a},S,T_{1},T_{2}\right\} $ and a
fermionic one $V_{1}=\left\{ Q_{\alpha }^{+},Q_{\alpha }^{-},R_{\alpha
}\right\} $ such that they satisfy 
\begin{equation}
\left[ V_{0},V_{0}\right] \subset V_{0}\,,\quad \left[ V_{0},V_{1}\right]
\subset V_{1}\,,\quad \left[ V_{1},V_{1}\right] \subset V_{0}\,. \label{subspace}
\end{equation}

In addition, a non-degenerate invariant supertrace on the Nappi-Witten superalgebra is given in terms of the following non-vanishing components:
\begin{eqnarray}
\left\langle G_{a}G_{b}\right\rangle &=&\delta_{ab}\,,\notag \\
\left\langle JS\right\rangle &=&-1\,,\notag \\
\left\langle T_1 T_2\right\rangle &=&1\,,\notag\\
\left\langle Q_{\alpha }^{-}Q_{\beta }^{-}\right\rangle &=&\left\langle Q_{\alpha }^{+}%
R_{\beta }\right\rangle\ =\ 2
C_{\alpha \beta }\,. \label{NWinv}
\end{eqnarray}
Then, the three-dimensional CS action based on the Nappi-Witten superalgebra can be constructed introducing the gauge connection one-form
\begin{equation}
A=\omega J+\omega^{a}G_{a}+sS+t_{1}T_{1}+t_{2}T_{2}+\bar{\psi}^{+}Q^{+}+\bar{\psi}^{-}Q^{-}+\bar{\rho}R
\end{equation}
and the previously defined invariant supertrace in the general expression of a CS action in three spacetime dimensions,
\begin{equation}
I_{\text{CS}}=\int\left\langle A \wedge dA+\frac{1}{3} A\wedge [A,A]\right\rangle \,.  \label{CS}
\end{equation}
The supersymmetric CS action invariant under the Nappi-Witten superalgebra \eqref{sNW} is given by\footnote{For the sake of simplicity, here and in the sequel we will omit writing the wedge product between forms and the spinor index $\alpha$ as well.}
\begin{equation}
I_{\text{NW}}=\int\omega_aR^{a}(\omega^{b})-2sR(\omega)+2t_{1}dt_{2}+2\bar{\psi}^{-}\nabla\psi^{-}+2\bar{\psi}^{+}\nabla\rho+2\bar{\rho}\nabla\psi^{+}\,,\label{CSactionNW}
\end{equation}
where
\begin{eqnarray}\label{RR}
R\left( \omega \right) &=&d\omega \,,  \notag \\
R^{a}\left( \omega ^{b}\right) &=&d\omega ^{a}+\epsilon ^{ac}\omega \omega
_{c}\,,
\end{eqnarray}
while the covariant derivatives of the spinor 1-form fields appearing in \eqref{CSactionNW} are
\begin{eqnarray}
\nabla \psi ^{+} &=&d\psi^{+}+\frac{1}{2}\omega \gamma _{0}\psi^{+}-\frac{1
}{2}t_{1}\gamma _{0}\psi^{+}\,,  \notag \\
\nabla \psi ^{-} &=&d\psi^{-}+\frac{1}{2}\omega \gamma _{0}\psi^{-}+\frac{1
}{2}\omega ^{a}\gamma _{a}\psi ^{+}+\frac{1}{2}t_{1}\gamma _{0}\psi^{-}\,, \notag  \\
\nabla \rho &=&d\rho +\frac{1}{2}\omega \gamma _{0}\rho +\frac{1}{2}\omega
^{a}\gamma _{a}\psi ^{-}+\frac{1}{2}s\gamma _{0}\psi^{+}-\frac{1}{2}
t_{2}\gamma _{0}\psi ^{+}-\frac{1}{2}t_{1}\gamma _{0}\rho\,. \label{cov}
\end{eqnarray}
Let us note that having a non-degenerate invariant trace correspond to the physical requirement that the CS supersymmetric action \eqref{CSactionNW} involves a kinematical term for each gauge field and the field equations reduce to the vanishing of all the curvatures of the model. In the present case, the curvatures for each gauge field are given by
\begin{eqnarray}
F\left( \omega \right) &=&R\left( \omega \right)+\frac{1}{2}\bar{\psi}^{+}\gamma^{0} \psi^{-} \,,  \notag \\
F^{a}\left( \omega ^{b}\right) &=&R^{a}\left( \omega ^{b}\right)+\bar{\psi}^{+}\gamma^{a} \psi^{-}\,, \notag\\
F\left(s\right)&=&ds+\frac{1}{2}\bar{\psi}^{-}\gamma^{0} \psi^{-}+\bar{\psi}^{+}\gamma^{0} \rho\,, \notag\\
F\left(t_{1}\right)&=&dt_{1}+\frac{1}{2}\bar{\psi}^{+}\gamma^{0}\psi^{+}\,, \notag \\
F\left(t_{2}\right)&=&dt_{2}-\frac{1}{2}\bar{\psi}^{-}\gamma^{0} \psi^{-}+\bar{\psi}^{+}\gamma^{0} \rho\,, \label{curvi}
\end{eqnarray}
along with \eqref{cov}. Then, the field equations coming from the variation of the CS action \eqref{CSactionNW} correspond, as it is expected, to the vanishing of all curvatures \eqref{cov} and \eqref{curvi}.

In the following sections, we will apply the $S$-expansion method \cite{Izaurieta:2006zz} to the super Nappi-Witten algebra in order to obtain different NR superalgebras. For our purpose, we will consider two different types of semigroup, that are $S_{E}^{\left( 2 N\right) }=\left\{ \lambda _{0},\lambda _{1},...,\lambda _{2N},\lambda _{2N+1}\right\} $ and $S_{\mathcal{M}}^{(2N)}=\lbrace \lambda_0,\lambda_1,\lambda_2,\ldots,\lambda_{2N-1},\lambda_{2N} \rbrace$. In both cases, for different values of $N$, known and new NR superalgebras will appear, corresponding to distinct supersymmetric extensions of known NR algebras.

One of the advantages of working with the $S$-expansion is that it provides us with the non-degenerate invariant tensors of the $S$-expanded (super)algebras, which are given in terms of the invariant tensors for the original (super)algebra. This turns out to be essential in the construction of NR CS (super)gravity actions.

\section{Generalized extended Bargmann supergravity theory and semigroup expansion method}\label{geb}

In this section, we shall derive different NR superalgebras whose bosonic sectors include as subalgebra the extended Bargmann algebra and its generalizations, depending on the case. The aforementioned NR superalgebras appear as the result of applying the $S$-expansion procedure to the Nappi-Witten superalgebra introduced in the previous section, considering $S_{E}^{\left( 2 N\right) }=\left\{ \lambda _{0},\lambda _{1},...,\lambda _{2N},\lambda _{2N+1}\right\} $ as the relevant semigroup.  We will show that all known and new NR superalgebras belong to a family of NR superalgebras, which we call as generalized extended Bargmann (GEB$
^{(N)}$) superalgebras. As we will see, they correspond to supersymmetric extensions of the NR counterparts of the $\mathfrak{B}_{N+2}$ algebras enlarged with $\left(N+1\right)$ $U_1$ generators \cite{Concha:2020sjt}. At the relativistic level, the $\mathfrak{B}_{N+2}$ algebras were first introduced in \cite{Edelstein:2006se,Izaurieta:2009hz} for obtaining standard General Relativity from Chern-Simons gravity.
Furthermore, as we have mentioned before, an important advantage of the $S$-expansion is that it allows to derive the non-vanishing components of an invariant tensor of the expanded (super)algebra. Then, we will exploit this powerful feature to derive the NR invariant supertraces and the CS actions invariant under the aforesaid expanded NR superalgebras.

\subsection{Extended Bargmann supergravity}\label{ebs}

The so-called extended Bargmann algebra \cite{Bergshoeff:2016lwr} can be obtained as the NR limit of the Poincaré algebra (corresponding to the $\mathfrak{B}_3$ algebra) enlarged with two $U_1$ generators. An alternative method has been proposed in \cite{Penafiel:2019czp} in which the extended Bargmann algebra can be derived as an $S$-expansion of the Nappi-Witten algebra. Here we show that a supersymmetric extension of the extended Bargmann algebra can be obtained by performing an $S$-expansion of the Nappi-Witten superalgebra (\ref{sNW}). To this end, we consider $S_{E}^{\left( 2\right) }=\left\{ \lambda _{0},\lambda _{1},\lambda_{2},\lambda _{3}\right\} $ as the relevant semigroup whose elements satisfy the following multiplication table:
\begin{equation}
\begin{tabular}{l|llll}
$\lambda _{3}$ & $\lambda _{3}$ & $\lambda _{3}$ & $\lambda _{3}$ & $\lambda
_{3}$ \\
$\lambda _{2}$ & $\lambda _{2}$ & $\lambda _{3}$ & $\lambda _{3}$ & $\lambda
_{3}$ \\
$\lambda _{1}$ & $\lambda _{1}$ & $\lambda _{2}$ & $\lambda _{3}$ & $\lambda
_{3}$ \\
$\lambda _{0}$ & $\lambda _{0}$ & $\lambda _{1}$ & $\lambda _{2}$ & $\lambda
_{3}$ \\ \hline
& $\lambda _{0}$ & $\lambda _{1}$ & $\lambda _{2}$ & $\lambda _{3}$%
\end{tabular}
\label{ML}
\end{equation}
Indeed, one can consider a resonant decomposition,
\begin{eqnarray}
S_{0} &=&\left\{ \lambda _{0},\lambda _{2},\lambda _{3}\right\} \,,  \notag
\\
S_{1} &=&\left\{ \lambda _{1},\lambda _{3}\right\} \,,  \label{SD}
\end{eqnarray}
with $\lambda _{3}=0_{S}$ being the zero element of the semigroup. Let us
note that the decomposition (\ref{SD}) is said to be resonant since it satisfies the same structure as the super Nappi-Witten subspaces,
\begin{equation}
S_{0}\cdot S_{0}\subset S_{0}\,,\qquad S_{0}\cdot S_{1}\subset S_{1}\,,\qquad
S_{1}\cdot S_{1}\subset S_{0}\,.
\end{equation}
Then, following the definitions of \cite{Izaurieta:2006zz}, after considering a resonant $S_{E}^{\left( 2\right) }$-expansion followed by a $0_{S}$-reduction to the
Nappi-Witten superalgebra (\ref{sNW}), we find an expanded superalgebra
spanned by the set of generators\footnote{Here and in the sequel, we denote the generators of the expanded algebra with a tilde symbol.} $\left\{ \tilde{J},\tilde{G}_{a},\tilde{S},\tilde{H},\tilde{P}_{a},\tilde{M},\tilde{Y}_1,\tilde{Y}_{2},\tilde{U}_1,\tilde{U}_2,\tilde{Q}_{\alpha }^{+},\tilde{Q}_{\alpha
}^{-},\tilde{R}_{\alpha }\right\} $ which are related to the super Nappi-Witten ones through the semigroup elements as follows:
\begin{eqnarray}
 \tilde{J}&=&\lambda _{0}J\,, \ \ \ \ \ \ \tilde{G}_{a}=\lambda _{0}G_{a}\,, \ \ \ \ \ \ \tilde{S}=\lambda _{0}S\,, \ \ \ \tilde{Y}_1=\lambda _{0}T_1\,, \ \ \ \ \tilde{U}_{1}=\lambda_{2}T_1\,, \notag\\
 \tilde{H}&=&\lambda _{2}J\,,\ \ \ \ \ \ \, \tilde{P}_{a}=\lambda _{2}G_{a}\,, \ \ \ \ \,\tilde{M}=\lambda _{2}S\,, \ \, \, \ \tilde{Y}_2=\lambda _{0}T_2\,, \ \ \ \ \tilde{U}_2=\lambda _{2}T_2\,, \notag\\
 \tilde{Q}_{\alpha }^{+}&=&\lambda _{1}Q_{\alpha }^{+}\,, \ \, \, \ \tilde{Q}
_{\alpha }^{-}=\lambda _{1}Q_{\alpha }^{-}\,,\ \, \, \ \tilde{R}_{\alpha
}=\lambda _{1}R_{\alpha }\,. \label{tab1}
\end{eqnarray}
Such expanded generators satisfy the following non-vanishing
(anti-)commutation relations:
\begin{eqnarray}
\left[ \tilde{J},\tilde{G}_{a}\right]  &=&\epsilon _{ab}\tilde{G}%
_{b}\,,\qquad \ \ \ \ \ \ \ \ \ \left[ \tilde{G}_{a},\tilde{G}_{b}\right] =-\epsilon
_{ab}\tilde{S}\,, \qquad \ \ \ \ \ \ \ \ \ \ \ \left[ \tilde{J}_{,}\tilde{P}_{a}\right] =\epsilon_{ab}\tilde{P}%
_{b} \notag \\
\left[ \tilde{H},\tilde{G}_{a}\right]  &=&\epsilon _{ab}\tilde{P}_{b}\,,\qquad \ \ \ \ \ \ \ \ \ \, \left[ \tilde{G}_{a},\tilde{P}_{b}\right] =-\epsilon_{ab}\tilde{M}\,,\qquad \ \ \ \ \ \ \ \  \left[ \tilde{J},\tilde{Q}_{\alpha }^{\pm }\right]  =-\frac{1}{2}\left( \gamma _{0}\right)
_{\alpha }^{\,\ \beta }
\tilde{Q}_{\beta }^{\pm }\,,\notag \\
\left[ \tilde{J},\tilde{R}%
_{\alpha }\right] &=&-\frac{1}{2}\left( \gamma _{0}\right)
_{\alpha }^{\,\ \beta }\tilde{R}_{\beta }\,, \ \ \  \left[ \tilde{G}_{a},\tilde{Q}_{\alpha }^{+}\right]  =-\frac{1}{2}\left( \gamma _{a}\right)
_{\alpha }^{\,\ \beta }\tilde{Q}_{\beta }^{-}\,, \ \ \ \ \left[ \tilde{G}_{a},\tilde{Q}
_{\alpha }^{-}\right] =-\frac{1}{2}\left( \gamma _{a}\right)
_{\alpha }^{\,\ \beta }\tilde{R}_{\beta }\,,  \notag \\
\left[ \tilde{S},\tilde{Q}_{\alpha }^{+}\right]  &=&-\frac{1}{2}\left( \gamma _{0}\right)
_{\alpha }^{\,\ \beta }\tilde{R}_{\beta }\,,  \ \ \ \, \left[
\tilde{Y}_{1},Q_{\alpha }^{\pm}\right] = \pm \frac{1}{2}\left( \gamma _{0}\right) _{\alpha
\beta }Q_{\beta }^{\pm}\,, \ \ \ \ \ \left[ \tilde{Y}_{1},R_{\alpha }\right] = \frac{1}{2}\left( \gamma _{0}\right)
_{\alpha \beta }R_{\beta }\,, \quad \ \ \ \ \ \notag \\
\left[ \tilde{Y}_{2},\tilde{Q}_{\alpha }^{+}%
\right] &=& \frac{1}{2}\left( \gamma _{0}\right) _{\alpha \beta }R_{\beta } \, \notag \\
\left\{ \tilde{Q}_{\alpha }^{+},\tilde{Q}
_{\beta }^{-}\right\} &=& -\left( \gamma ^{a}C\right) _{\alpha \beta }\tilde{P}
_{a}\,,  \notag \\
\left\{ \tilde{Q}_{\alpha }^{+},\tilde{Q}_{\beta }^{+}\right\}  &=&-\left(
\gamma ^{0}C\right) _{\alpha \beta }\tilde{H} -\left(
\gamma ^{0}C\right) _{\alpha \beta }\tilde{U}_1\,, \notag \\
\left\{ \tilde{Q}_{\alpha }^{-},\tilde{Q}_{\beta }^{-}\right\} &=& -\left( \gamma ^{0}C\right)
_{\alpha \beta }\tilde{M} + \left(\gamma ^{0}C\right) _{\alpha \beta }\tilde{U}_2 \,,  \notag \\
\left\{ \tilde{Q}_{\alpha }^{+},\tilde{R}_{\beta }\right\}  &=& - \left(
C\gamma ^{0}\right) _{\alpha \beta }\tilde{M} -\left(\gamma ^{0}C\right) _{\alpha \beta }\tilde{U}_2 \,. \label{sEB}
\end{eqnarray}
This superalgebra corresponds to an extension of the extended Bargmann superalgebra introduced in \cite{Bergshoeff:2016lwr}.\footnote{Note that we have some differences in signs with respect to the commutation relations of \cite{Bergshoeff:2016lwr}, due to different conventions, but this can be solved by just setting $\epsilon_{ab} \rightarrow - \epsilon_{ab}$; also, let us recall that $\gamma_0 = - \gamma^0$.} In fact, by considering $\tilde{Y}_1=\tilde{Y}_2=\tilde{U}_1= \tilde{U}_2=0$, one can exactly reproduce the extended Bargmann superalgebra of \cite{Bergshoeff:2016lwr}. Let us note that the additional bosonic generators $\tilde{Y}_1,\,\tilde{Y}_2,\,\tilde{U}_1$, and $\tilde{U}_2$ appear as a consequence of the expansion of the Nappi-Witten generators $T_1$ and $T_2$. Their presence is essential to get a non-degenerate invariant tensor.

The invariant tensor for the extended Bargmann superalgebra can be obtained by applying the $S$-expansion method to the invariant tensor of the Nappi-Witten superalgebra, given in \eqref{NWinv}.  In this way, one finds that the non-vanishing components of a non-degenerate invariant tensor for the extended Bargmann superalgebra are given by
\begin{eqnarray}
\left\langle \tilde{G}_a \tilde{G}_b \right\rangle &=& \alpha_0
\delta_{ab} \,, \qquad \ \qquad \qquad  \  \ \ \ \left\langle \tilde{J} \tilde{S} \right\rangle = -\alpha_0 \,,\notag \\
\left\langle \tilde{G}_a \tilde{P}_b \right\rangle &=& \alpha_1
\delta_{ab} \,, \qquad \qquad \qquad \ \ \ \left\langle \tilde{Y}_{1}\tilde{Y}_{2}\right\rangle =\alpha_{0}\,,\notag
\\
\left\langle \tilde{J} \tilde{M} \right\rangle &=& \left\langle \tilde{H} \tilde{S} \right\rangle \, = \,
-\alpha_{1}\,,\qquad \  \ \left\langle \tilde{Y}_{1}\tilde{U}_{2}\right\rangle \,=\,\left\langle \tilde{U}_{1}\tilde{Y}_{2}\right\rangle\, = \,
\alpha_{1} \,, \notag\\
\left\langle \tilde{Q}_{\alpha }^{-}\tilde{Q}_{\beta }^{-}\right\rangle &=&\left\langle \tilde{Q}_{\alpha }^{+}
\tilde{R}_{\beta }\right\rangle=2
\alpha_{1}C_{\alpha \beta }\,,
\label{invt1}
\end{eqnarray}
where the $\alpha$'s are arbitrary constants and appear as a consequence of the $S$-expansion procedure. Then, the three-dimensional CS action invariant under the superalgebra \eqref{sEB} can be directly constructed by introducing the gauge connection one-form
\begin{align}
A=\ &\omega \tilde{J}+h\tilde{H}+\omega^{a}\tilde{G}_{a}+e^{a}\tilde{P}_{a}+s\tilde{S}+m\tilde{M}+y_{1}\tilde{Y}_{1}+y_{2}\tilde{Y}_{2}+u_{1}\tilde{U}_{1}+u_{2}\tilde{U}_{2}\notag\\
&+\bar{\psi}^{+}\tilde{Q}^{+}+\bar{\psi}^{-}\tilde{Q}^{-}+\bar{\rho}\tilde{R}
\end{align}
and the invariant tensor \eqref{invt1} in the general expression for a CS action \eqref{CS}. The NR CS supergravity action reads
\begin{equation}\label{CSEB}
    I_{\text{EB}}=\alpha_{0}I_{0}+\alpha_{1}I_{1} \,,
\end{equation}
where
\begin{eqnarray}
I_{0}&=&\int\omega _{a}R^{a}(\omega ^{b})-2sR\left(
\omega \right) +2y_{1}dy_{2}\,, \notag\\
I_{1}&=&\int\left.2e_{a}R^{a}(\omega ^{b})-2mR(\omega )-2\tau R(s)+2y_{1}du_{2}+2u_{1}dy_{2}\right.   \notag\\
&&\left. +2\bar{\psi}^{+}\nabla \rho+ 2\bar{\rho}\nabla \psi ^{+}+2\bar{\psi}^{-}\nabla \psi^{-}\right.\,,\label{I0I1}
\end{eqnarray}
and 
\begin{equation}
R(s)=ds +\frac{1}{2}\epsilon^{ac}\omega_a\omega_c\,,\label{RS}
\end{equation}
while $R(\omega)$ and $R^{a}(\omega^{b})$ are given by \eqref{RR}. Besides, the covariant derivatives of the spinor 1-form fields read
\begin{eqnarray}
\nabla \psi ^{+} &=&d\psi^{+}+\frac{1}{2}\omega \gamma _{0}\psi^{+}-\frac{1
}{2}y_{1}\gamma _{0}\psi^{+}\,,  \notag \\
\nabla \psi ^{-} &=&d\psi^{-}+\frac{1}{2}\omega \gamma _{0}\psi^{-}+\frac{1
}{2}\omega ^{a}\gamma _{a}\psi ^{+}+\frac{1}{2}y_{1}\gamma _{0}\psi^{-}\,, \notag  \\
\nabla \rho &=&d\rho +\frac{1}{2}\omega \gamma _{0}\rho +\frac{1}{2}\omega
^{a}\gamma _{a}\psi ^{-}+\frac{1}{2}s\gamma _{0}\psi^{+}-\frac{1}{2}
y_{2}\gamma _{0}\psi ^{+}-\frac{1}{2}y_{1}\gamma _{0}\rho\,.  \label{fermcurvEB}
\end{eqnarray}

The CS action \eqref{CSEB} describes an extension of the so-called extended Bargmann supergravity theory \cite{Bergshoeff:2016lwr}. Indeed, the CS action $I_1$ corresponds to the extended Bargmann supergravity action introduced in \cite{Bergshoeff:2016lwr}, endowed with some additional terms involving the extra bosonic 1-form fields $y_1$, $y_2$, $u_1$, and $u_2$. Furthermore, the term along $\alpha_0$ corresponds to a NR exotic Lagrangian. The extra bosonic field content is related to the additional bosonic generators which allow to define the non-degenerate invariant tensor \eqref{invt1}. The non-degeneracy of the invariant tensor implies that the equations of motion are given by the vanishing of the curvature two-forms of the model, which are given by \eqref{fermcurvEB} and
\begin{eqnarray}
F\left( \omega \right) &=&R(\omega) \,,  \qquad \qquad \qquad \qquad \qquad \qquad \quad \,
F^{a}\left( \omega ^{b}\right) =R^{a}(\omega^{b})\,, \notag\\
F\left(s\right)&=&R(s)\,, \qquad \qquad \qquad \qquad \qquad \quad \quad \quad \, F\left( \tau \right) =d\left( \tau \right)+\frac{1}{2}\bar{\psi}^{+}\gamma^{0} \psi^{-} \,,  \notag \\
F^{a}\left( e ^{b}\right) &=&de^{a}+\epsilon^{ac}\omega e_c+\epsilon^{ac}\tau \omega_c+\bar{\psi}^{+}\gamma^{a} \psi^{-}\,, \quad 
F\left(m\right)=dm+\epsilon^{ac}\omega_a e_c+\frac{1}{2}\bar{\psi}^{-}\gamma^{0} \psi^{-}+\bar{\psi}^{+}\gamma^{0} \rho\,, \notag\\
F\left(y_{1}\right)&=&dy_{1}\,, \qquad \qquad \qquad \qquad \qquad \qquad \quad \ \
F\left(y_{2}\right)=dy_{2}\,, \notag \\
F\left(u_{1}\right)&=&du_{1}+\frac{1}{2}\bar{\psi}^{+}\gamma^{0}\psi^{+}\,, \qquad \qquad \qquad \quad \ \ \ 
F\left(u_{2}\right)=du_{2}-\frac{1}{2}\bar{\psi}^{-}\gamma^{0} \psi^{-}+\bar{\psi}^{+}\gamma^{0} \rho\,. \label{curvi2}
\end{eqnarray}
Let us note that the present NR theory can be seen as the most general three-dimensional extended Bargmann supergravity theory containing both exotic and standard sectors. Nonetheless, the formulation of a NR supergravity theory is not unique and can be generalized beyond the extended Bargmann one.

\subsection{Maxwellian extended Bargmann supergravity}\label{mebs}

A Maxwellian version of the extended Bargmann algebra was recently presented in \cite{Aviles:2018jzw}. It was denoted as
MEB algebra and it was obtained as a NR limit of the Maxwell algebra (also called as $\mathfrak{B}_4 $ algebra) enlarged with three $U_1$ generators. At the relativistic level, the Maxwell symmetry appears in the description of a particle in a Minkowski spacetime in the presence of an electromagnetic field background \cite{Schrader:1972zd,Bacry:1970ye,Gomis:2017cmt}. Subsequently, a supersymmetric extension of the MEB algebra was introduced in \cite{Concha:2019mxx}. In order to construct a well-defined supergravity theory in this context it was necessary to construct by hand the aforementioned supersymmetric extension. Now, we are going to show that the MEB superalgebra and the corresponding NR supergravity theory can be derived by means of the $S$-expansion method. Indeed, let us consider the $S_{E}^{\left(
4\right) }$-expansion of the Nappi-Witten superalgebra (\ref{sNW}). The
elements of the $S_{E}^{\left( 4\right) }$ semigroup satisfy
\begin{equation}
\begin{tabular}{l|llllll}
$\lambda _{5}$ & $\lambda _{5}$ & $\lambda _{5}$ & $\lambda _{5}$ & $\lambda
_{5}$ & $\lambda _{5}$ & $\lambda _{5}$ \\
$\lambda _{4}$ & $\lambda _{4}$ & $\lambda _{5}$ & $\lambda _{5}$ & $\lambda
_{5}$ & $\lambda _{5}$ & $\lambda _{5}$ \\
$\lambda _{3}$ & $\lambda _{3}$ & $\lambda _{4}$ & $\lambda _{5}$ & $\lambda
_{5}$ & $\lambda _{5}$ & $\lambda _{5}$ \\
$\lambda _{2}$ & $\lambda _{2}$ & $\lambda _{3}$ & $\lambda _{4}$ & $\lambda
_{5}$ & $\lambda _{5}$ & $\lambda _{5}$ \\
$\lambda _{1}$ & $\lambda _{1}$ & $\lambda _{2}$ & $\lambda _{3}$ & $\lambda
_{4}$ & $\lambda _{5}$ & $\lambda _{5}$ \\
$\lambda _{0}$ & $\lambda _{0}$ & $\lambda _{1}$ & $\lambda _{2}$ & $\lambda
_{3}$ & $\lambda _{4}$ & $\lambda _{5}$ \\ \hline
& $\lambda _{0}$ & $\lambda _{1}$ & $\lambda _{2}$ & $\lambda _{3}$ & $%
\lambda _{4}$ & $\lambda _{5}$%
\end{tabular}
\label{ML3}
\end{equation}%
where $\lambda _{5}=0_{S}$ is the zero element of the semigroup. Let $%
S_{E}^{\left( 4\right) }=S_{0}\cup S_{1}$ be a resonant decomposition with%
\begin{eqnarray}
S_{0} &=&\left\{ \lambda _{0},\lambda _{2},\lambda _{4},\lambda _{5}\right\}
\,,  \notag \\
S_{1} &=&\left\{ \lambda _{1},\lambda _{3},\lambda _{5}\right\} \,.
\end{eqnarray}%
After considering a resonant $S_{E}^{\left(
4\right) }$-expansion followed by a $0_{S}$-reduction of the Nappi-Witten superalgebra, we find an expanded
superalgebra spanned by the generators
\begin{equation*}
\left\{ \tilde{J},\tilde{G}_{a},\tilde{S},\tilde{H},\tilde{P}_{a},\tilde{M},%
\tilde{Z},\tilde{Z}_{a},\tilde{T},\tilde{Y}_{1},\tilde{Y}_{2},\tilde{U}_{1},%
\tilde{U}_{2},\tilde{B}_{1},\tilde{B}_{2},\tilde{Q}_{\alpha }^{+},\tilde{Q}%
_{\alpha }^{-},\tilde{R}_{\alpha },\tilde{\Sigma}_{\alpha }^{+},\tilde{\Sigma%
}_{\alpha }^{-},\tilde{W}_{\alpha }\right\}\,,
\end{equation*}
which are related to the super Nappi-Witten ones through
\begin{eqnarray}
 \tilde{J}&=&\lambda _{0}J\,, \qquad \tilde{G}_{a}=\lambda _{0}G_{a}\,,\qquad \ \tilde{S}=\lambda _{0}S\,, \qquad
\tilde{Y}_{1}=\lambda _{0}T_{1}\,, \qquad \, \tilde{Y}_{2}=\lambda _{0}T_{2}\,, \notag \\
\tilde{H}&=&\lambda _{2}J\,,\qquad \tilde{P}_{a}=\lambda _{2}G_{a}\,,\qquad \tilde{M}=\lambda _{2}S\,, \qquad
\tilde{U}_{1}=\lambda _{2}T_{1}\,, \qquad \tilde{U}_{2}=\lambda _{2}T_{2}\,, \notag \\
\tilde{Z}&=&\lambda _{4}J\,,\qquad \tilde{Z}_{a}=\lambda _{4}G_{a}\,,\, \qquad \tilde{T}=\lambda _{4}S\,, \qquad
\tilde{B}_{1}=\lambda _{4}T_{1}\,, \qquad \tilde{B}_{2}=\lambda _{4}T_{2}\,, \notag \\
\tilde{Q}_{\alpha }^{+}&=& \lambda _{1}Q_{\alpha }^{+}\,, \, \, \, \ \tilde{Q}_{\alpha }^{-}=\lambda _{1}Q_{\alpha }^{-}\,, \quad \ \tilde{\Sigma}
_{\alpha }^{+}=\lambda _{3}Q_{\alpha }^{+}\,,\quad \, \tilde{\Sigma}%
_{\alpha }^{-}=\lambda _{3}Q_{\alpha }^{-}\,,\quad \ 
 \tilde{R}_{\alpha
}=\lambda _{1}R_{\alpha }\,, \notag \\
 \tilde{W}_{\alpha
}&=&\lambda _{3}R_{\alpha }\,.
\end{eqnarray}

Using the multiplication law (\ref{ML3}) and the original (anti-)commutation relations of the Nappi-Witten superalgebra (\ref{sNW}), one can show that the expanded generators satisfy an expanded NR superalgebra whose non-trivial (anti-)commutation relations are given by \eqref{sEB} along with
\begin{eqnarray}
\left[ \tilde{P}_{a},\tilde{P}_{b}\right]  &=&-\epsilon _{ab}\tilde{T}
\,,\qquad \ \ \ \ \ \ \ \ \ \left[ \tilde{G}_{a},\tilde{Z}_{b}\right] =-\epsilon _{ab}\tilde{T}\,,\qquad \ \ \ \ \ \ \ \ \ \ \ \left[ \tilde{H},\tilde{P}_{a}\right]  =\epsilon _{ab}\tilde{Z}
_{b}\,, \notag \\
\left[ \tilde{J},\tilde{Z}_{a}\right] &=&\epsilon _{ab}
\tilde{Z}_{b}\,, \qquad \ \ \ \ \ \ \ \ \ \ \ 
\left[ \tilde{Z},\tilde{G}_{a}\right] =\epsilon _{ab}\tilde{Z}
_{b}\,,  \qquad \ \ \ \ \ \ \ \ \ \ \ 
\left[ \tilde{U}_{2},\tilde{Q}_{\alpha }^{+}\right]  =\frac{1}{2}
\left( \gamma _{0}\right) _{\alpha \beta }\tilde{W}_{\beta }\,,\notag \\
 \left[ \tilde{J},\tilde{\Sigma}_{\alpha }^{\pm }\right] &=&-
\frac{1}{2}\left( \gamma _{0}\right) _{\alpha }^{\text{ }\beta }\tilde{\Sigma%
}_{\beta }^{\pm }\,,\ \ \ \ \ \ \ \left[ \tilde{S},\tilde{\Sigma}_{\alpha }^{+}\right] =-\frac{1}{2}
\left( \gamma _{0}\right) _{\alpha }^{\text{ }\beta }\tilde{W}_{\beta }\,,\qquad  \left[ \tilde{Y}_{1},\tilde{W}_{\alpha }\right] =\frac{1%
}{2}\left( \gamma _{0}\right) _{\alpha \beta }\tilde{W}_{\beta }\,,  \notag \\
\left[ \tilde{H},\tilde{Q}_{\alpha }^{\pm }\right]  &=&-\frac{1}{2}\left(
\gamma _{0}\right) _{\alpha }^{\text{ }\beta }\tilde{\Sigma}_{\beta }^{\pm
}\,,\quad \ \ \left[ \tilde{P}_{a},\tilde{Q}_{\alpha }^{+}\right] =-\frac{1}{2%
}\left( \gamma _{a}\right) _{\alpha }^{\text{ }\beta }\tilde{\Sigma}_{\beta
}^{-}\,,\qquad \ \left[ \tilde{Y}_{1},\tilde{\Sigma}_{\alpha }^{+}\right]  =\frac{1}{2}%
\left( \gamma _{0}\right) _{\alpha \beta }\tilde{\Sigma}_{\beta
}^{+}\,,  \notag \\
\left[ \tilde{G}_{a},\tilde{\Sigma}_{\alpha }^{+}\right]  &=&-\frac{1}{2}%
\left( \gamma _{a}\right) _{\alpha }^{\text{ }\beta }\tilde{\Sigma}_{\beta
}^{-}\,, \quad \ \ \left[ \tilde{G}_{a},\tilde{\Sigma}_{\alpha }^{-}%
\right] =-\frac{1}{2}\left( \gamma _{a}\right) _{\alpha }^{\text{ }\beta }%
\tilde{W}_{\beta }\,, \qquad\left[ \tilde{Y}_{1},\tilde{\Sigma}_{\alpha }^{-}\right]  = - \frac{1}{2}%
\left( \gamma _{0}\right) _{\alpha \beta }\tilde{\Sigma}_{\beta
}^{-}\,, \notag \\
\left[ P_{a},\tilde{Q}_{\alpha }^{-}\right]  &=&-\frac{1}{2}\left( \gamma
_{a}\right) _{\alpha }^{\text{ }\beta }\tilde{W}_{\beta }\,,\qquad \ %
\left[ \tilde{J},\tilde{R}_{\alpha }\right] =-\frac{1}{2}\left( \gamma
_{0}\right) _{\alpha }^{\text{ }\beta }\tilde{R}_{\beta }\,,\qquad \left[ \tilde{U}_{1},\tilde{Q}_{\alpha }^{+}\right]  =\frac{1}{2}\left(
\gamma _{0}\right) _{\alpha \beta }\tilde{\Sigma}_{\beta }^{+}\,,\notag\\
\left[ \tilde{J},\tilde{W}_{\alpha }\right]  &=&-\frac{1}{2}\left( \gamma
_{0}\right) _{\alpha }^{\text{ }\beta }\tilde{W}_{\beta }\,,\qquad %
\left[ \tilde{H},\tilde{R}_{\alpha }\right] =-\frac{1}{2}\left( \gamma
_{0}\right) _{\alpha }^{\text{ }\beta }\tilde{W}_{\beta }\,, \qquad \left[ \tilde{U}_{1},\tilde{Q}_{\alpha }^{-}\right] = - \frac{1}{2}\left(
\gamma _{0}\right) _{\alpha \beta }\tilde{\Sigma}_{\beta }^{-}\,, \notag \\
\left[ \tilde{M},\tilde{Q}_{\alpha }^{+}\right]  &=&-\frac{1}{2}\left(
\gamma _{0}\right) _{\alpha }^{\text{ }\beta }\tilde{W}_{\beta }\,, \qquad  \left[ \tilde{Y}_{2},\tilde{\Sigma}_{\alpha }^{+}\right]
=\frac{1}{2}\left( \gamma _{0}\right) _{\alpha \beta }\tilde{W}_{\beta }\,,\qquad \ \left[ \tilde{U}_{1},\tilde{R}_{\alpha }\right] =\frac{1}{2}\left( \gamma
_{0}\right) _{\alpha \beta }\tilde{W}_{\beta }\,,\notag\\
\left\{ \tilde{Q}_{\alpha }^{-},\tilde{\Sigma}_{\beta }^{-}\right\}
&=&-\left( \gamma ^{0}C\right) _{\alpha \beta }\tilde{T} {+} \left( \gamma
^{0}C\right) _{\alpha \beta }\tilde{B}_{2}\,,  \notag \\
\left\{ \tilde{Q}_{\alpha }^{\pm },\tilde{\Sigma}_{\beta }^{\mp }\right\}
&=&-\left( \gamma ^{a}C\right) _{\alpha \beta }\tilde{Z}_{a}\,,\notag \\
\left\{ \tilde{Q}_{\alpha }^{+},\tilde{\Sigma}_{\beta }^{+}\right\}
&=&-\left( \gamma ^{0}C\right) _{\alpha \beta }\tilde{Z}-\left( \gamma
^{0}C\right) _{\alpha \beta }\tilde{B}_{1}\,,  \notag \\
\left\{ \tilde{Q}_{\alpha }^{+},\tilde{W}_{\beta }\right\}  &=&\left\{ \tilde{\Sigma}_{\alpha }^{+},\tilde{R}_{\beta }\right\} =-\left(
\gamma ^{0}C\right) _{\alpha \beta }\tilde{T}-\left( \gamma ^{0}C\right)
_{\alpha \beta }\tilde{B}_{2}\,. \label{sMEB3}
\end{eqnarray}
The superalgebra just obtained corresponds to the Maxwellian extended Bargmann superalgebra first presented in \cite{Concha:2019mxx}. The MEB superalgebra is characterized by the presence of three additional fermionic generators, besides $\tilde{Q}^{\pm}_{\alpha}$ and $\tilde{R}_{\alpha}$, namely $\tilde{\Sigma}_{\alpha}^{+},\tilde{\Sigma}_{\alpha}^{-}$, and $\tilde{W}_{\alpha}$. Furthermore, unlike the extended Bargmann superalgebra, $\tilde{U_1}$ and $\tilde{U_2}$ are no longer central charges but act non-trivially on the fermionic generators $\tilde{Q}_{\alpha}^{\pm}$ and $\tilde{R}_{\alpha}$. The non-vanishing components of an invariant supertrace for the MEB superalgebra are obtained from the Nappi-Witten ones through the $S$-expansion method. These components are given by \eqref{invt1} together with
\begin{eqnarray}
\left\langle \tilde{P}_{a}\tilde{P}_{b}\right\rangle &=&\left\langle \tilde{G%
}_{a}\tilde{Z}_{b}\right\rangle \ =\ \alpha_{2}\delta _{ab}\,,\notag \\
\left\langle \tilde{J}\tilde{T}\right\rangle &=&\left\langle \tilde{H}\tilde{%
M}\right\rangle \ =\ \left\langle \tilde{S}\tilde{Z}\right\rangle \ = \ -\alpha
_{2}\,,\notag \\
\left\langle \tilde{Y}_{1}\tilde{B}_{2}\right\rangle &=&\left\langle \tilde{U}_{1}\tilde{U}_{2}\right\rangle
\,=\,\left\langle \tilde{B}_{1}\tilde{Y}_{2}\right\rangle \ =\ \alpha_2 \,,\notag \\
\left\langle \tilde{Q}_{\alpha }^{-}\tilde{\Sigma}_{\beta }^{-}\right\rangle
&=&\left\langle \tilde{\Sigma}%
_{\alpha }^{+}\tilde{R}_{\beta }\right\rangle \,=\,\left\langle \tilde{Q}%
_{\alpha }^{+}\tilde{W}_{\beta }\right\rangle \ =\ 2\alpha_2 C_{\alpha \beta } \,, \label{invMB}
\end{eqnarray}%
where $\alpha_2$ is an arbitrary constant. The NR CS supergravity action invariant under the MEB superalgebra, which was presented in \cite{Concha:2019mxx}, is obtained by considering the following gauge connection one-form:
\begin{eqnarray}
A &=&\omega \tilde{J}+\omega ^{a}\tilde{G}_{a}+\tau \tilde{H}+e^{a}
\tilde{P}_{a}+k\tilde{Z}+k^{a}\tilde{Z}_{a}+m\tilde{M}+s\tilde{S}+t\tilde{T}
\notag \\
&&+y_{1}\tilde{Y}_{1}+y_{2}\tilde{Y}_{2}+b_1\tilde{B}_{1}+b_{2}\tilde{B}
_{2}+u_{1}\tilde{U}_{1}+u_{2}\tilde{U}_{2}  \notag \\
&&+{\psi }^{+}\tilde{Q}^{+}+{\psi }^{-}\tilde{Q}^{-}+{\xi }^{+}\tilde{\Sigma}
^{+}+{\xi }^{-}\tilde{\Sigma}^{-}+{\rho }\tilde{R}+{\chi }\tilde{W} \label{oneformMEB}
\end{eqnarray}
and the non-vanishing components of the invariant tensor \eqref{invt1} and \eqref{invMB} in the CS expression \eqref{CS}. It reads, up to boundary terms, as follows:
\begin{equation}\label{CSMEB}
    I_{\text{MEB}}=\alpha_{0}I_{0}+\alpha_{1}I_{1}+\alpha_{2}I_{2} \,,
\end{equation}
where $I_0$ and $I_1$ are given in \eqref{I0I1}, while the term along $\alpha_2$ reads
\begin{eqnarray}
I_{2} &=&\int \left. e_{a}R^{a}\left( e^{b}\right)+k_{a}R^{a}\left( \omega ^{b}\right)+\omega _{a}R^{a}\left( k^{b}\right)-2sR\left( k\right)\right.  \notag \\
&&-\left.\left. 2mR\left(
\tau \right) -2tR\left( \omega \right)+2y_{1}db_{2}+2u_{1}du_{2}+2y_{2}db_{1}\right.\right.  \notag \\
&&+\left. 2\bar{\psi}^{-}\nabla \xi ^{-}+2
\bar{\xi}^{-}\nabla \psi ^{-}+2\bar{\psi}^{+}\nabla \chi +2\bar{\chi}\nabla
\psi ^{+}+2\bar{\xi}^{+}\nabla \rho +2\bar{\rho}\nabla \xi ^{+}\right.\,.
\label{MEBI2}
\end{eqnarray}
Here, $R(\omega)$ and $R^{a}(\omega^{b})$ were defined in \eqref{RR}, while 
\begin{eqnarray}
R\left( \tau \right) &=&d\tau \,,  \notag \\
R^{a}\left( e^{b}\right) &=&de^{a}+\epsilon ^{ac}\omega e_{c}+\epsilon
^{ac}\tau \omega _{c}\,,  \notag \\
R\left( k\right) &=&dk\,,  \notag \\
R^{a}\left( k^{b}\right) &=&dk^{a}+\epsilon ^{ac}\omega k_{c}+\epsilon
^{ac}\tau e_{c}+\epsilon ^{ac}k\omega _{c}\,. \label{curvi21}
\end{eqnarray}%
On the other hand, the covariant
derivatives of the spinor $1$-form fields appearing in $I_{2}$ are given by
\begin{eqnarray}
\nabla \xi ^{+} &=&d\xi ^{+}+\frac{1}{2}\omega \gamma _{0}\xi ^{+}+\frac{1}{2%
}\tau \gamma _{0}\psi ^{+}-\frac{1}{2}y_{1}\gamma _{0}\xi ^{+}-\frac{1}{2}%
u_{1}\gamma _{0}\psi ^{+}\,,  \notag \\
\nabla \xi ^{-} &=&d\xi ^{-}+\frac{1}{2}\omega \gamma _{0}\xi ^{-}+\frac{1}{2%
}\tau \gamma _{0}\psi ^{-}+\frac{1}{2}e^{a}\gamma _{a}\psi ^{+}+\frac{1}{2}%
\omega ^{a}\gamma _{a}\xi ^{+}+\frac{1}{2}y_{1}\gamma _{0}\xi ^{-}+\frac{1}{2%
}u_{1}\gamma _{0}\psi ^{-}\,,
\notag \\
\nabla \chi &=&d\chi +\frac{1}{2}\omega \gamma _{0}\chi +\frac{1}{2}\omega
^{a}\gamma _{a}\xi ^{-}+\frac{1}{2}e^{a}\gamma _{a}\psi ^{-}+\frac{1}{2}\tau
\gamma _{0}\rho +\frac{1}{2}s\gamma _{0}\xi ^{+}+\frac{1}{2}m\gamma _{0}\psi
^{+}  \notag \\
&&-\frac{1}{2}y_{2}\gamma _{0}\xi ^{+}-\frac{1}{2}y_{1}\gamma _{0}\chi -%
\frac{1}{2}u_{2}\gamma _{0}\psi ^{+}-\frac{1}{2}u_{1}\gamma _{0}\rho\,, \label{fermcurv}
\end{eqnarray}
along with \eqref{fermcurvEB}. The CS action \eqref{CSMEB} describes the Maxwellian extended Barmann supergravity theory first presented in \cite{Concha:2019mxx}. As we can see, the $S$-expansion of the super Nappi-Witten algebra \eqref{sNW} with the $S_{E}^{\left(4\right) }$ semigroup adds a new sector to the action with respect to the case of the extended Bargmann supergravity theory previously studied. This new sector along the arbitrary constant $\alpha_2$ corresponds to the CS action for a new NR Maxwell superalgebra, whose bosonic part is the MEB gravity presented in \cite{Aviles:2018jzw}, supplemented with some bosonic 1-form fields. Let us note that the extended Bargmann supergravity action \eqref{I0I1} appears as a particular subcase along $\alpha_0$ and $\alpha_1$. The equations of motion of the theory are given by the vanishing of the curvature two-forms, which, in the case under analysis, are given by \eqref{fermcurvEB}, \eqref{curvi2}, \eqref{fermcurv}, and
\begin{eqnarray}
F\left(k\right)&=&R\left(k\right)+\bar{\psi}^{+}\gamma^{0}\xi^{+}\,, \notag \\
F^{a}\left( k^{b}\right) &=&R^{a}\left(k^{b}\right)+\bar{\psi}^{+}\gamma^{a} \xi^{-}+\bar{\psi}^{-}\gamma^{a} \xi^{+}\,, \notag \\
F\left(t\right)&=&dt+\epsilon^{ac}\omega_a k_c+\frac{1}{2}\epsilon^{ac}e_a e_c+\bar{\psi}^{-}\gamma^{0} \xi^{-}+\bar{\psi}^{+}\gamma^{0} \chi+\bar{\xi}^{+}\gamma^{0} \rho\,, \notag\\
F\left(b_{1}\right)&=&db_{1}+\bar{\psi}^{+}\gamma^{0}\xi^{+}\,, \notag \\
F\left(b_{2}\right)&=&db_{2}-\bar{\psi}^{-}\gamma^{0} \xi^{-}+\bar{\psi}^{+}\gamma^{0} \chi+\bar{\xi}^{+}\gamma^{0} \rho\,. \label{curvi3}
\end{eqnarray}
This is indeed expected for a well-defined and consistent CS (super)gravity model.

\subsection{Generalized Maxwellian extended Bargmann supergravity}

A generalization of the Maxwellian extended Bargmann algebra was introduced very recently in \cite{Concha:2020sjt}. The aforesaid algebra was obtained by considering a NR limit of a generalized Maxwell algebra (also denoted as $\mathfrak{B}_{5}$) defined in three spacetime dimensions. Here, we will show that a supersymmetric extension of the generalized Maxwellian extended Bargmann
(GMEB) algebra can be obtained considering an $S_{E}^{\left(
6\right) }$-expansion of the Nappi-Witten superalgebra (\ref{sNW}). Furthermore, the $S$-expansion method will allow us to construct the supergravity theory invariant under the GMEB superalgebra, as it provides with the non-degenerate invariant supertrace.

The elements of the $S_{E}^{\left( 6\right) }$ semigroup satisfy the multiplication law
\begin{equation}
\lambda _{\alpha }\lambda _{\beta }=\left\{
\begin{array}{lcl}
\lambda _{\alpha +\beta }\,\,\,\, & \mathrm{if}\,\,\,\,\alpha +\beta \leq 6\,, &
\\
\lambda _{7}\,\, & \mathrm{if}\,\,\,\,\alpha +\beta > 6\,, &
\end{array}
\right.  \label{SE6}
\end{equation}
with $\lambda _{7}=0_{S}$ being the zero element of the semigroup.
 Let $S_{E}^{\left( 6\right) }=S_{0}\cup S_{1}$ be a resonant decomposition with
\begin{eqnarray}
S_{0} &=&\left\{ \lambda _{0},\lambda _{2},\lambda _{4},\lambda _{6}, \lambda_{7}\right\}
\,,  \notag \\
S_{1} &=&\left\{ \lambda _{1},\lambda _{3},\lambda _{5}, \lambda_{7}\right\} \,.
\end{eqnarray}%
After applying a resonant $S_{E}^{\left(
6\right) }$-expansion and a $0_{S}$-reduction to the Nappi-Witten superalgebra, we find an expanded superalgebra spanned by the bosonic generators
\begin{equation*}
\left\{ \tilde{J},\tilde{G}_{a},\tilde{S},\tilde{H},\tilde{P}_{a},\tilde{M},%
\tilde{Z},\tilde{Z}_{a},\tilde{N},\tilde{N}_{a},\tilde{T},\tilde{V},\tilde{Y}_{1},\tilde{Y}_{2},\tilde{U}_{1},%
\tilde{U}_{2},\tilde{B}_{1},\tilde{B}_{2},\tilde{C}_{1},\tilde{C}_{2}\right\}
\end{equation*}
along with the fermionic charges
\begin{equation*}
\left\{\tilde{Q}_{\alpha }^{+},\tilde{Q}_{\alpha}^{-},\tilde{R}_{\alpha },\tilde{\Sigma}_{\alpha }^{+},\tilde{\Sigma
}_{\alpha }^{-},\tilde{W}_{\alpha },\tilde{\Xi}_{\alpha }^{+},\tilde{\Xi
}_{\alpha }^{-},\tilde{K}_{\alpha }\right\} \,,
\end{equation*}
which are related to the super Nappi-Witten ones through
\begin{eqnarray}
 \tilde{J}&=&\lambda _{0}J\,, \qquad \tilde{G}_{a}=\lambda _{0}G_{a}\,,\qquad \ \tilde{S}=\lambda _{0}S\,, \qquad
\tilde{Y}_{1}=\lambda _{0}T_{1}\,, \qquad \, \tilde{Y}_{2}=\lambda _{0}T_{2}\,, \notag \\
\tilde{H}&=&\lambda _{2}J\,,\qquad \tilde{P}_{a}=\lambda _{2}G_{a}\,,\qquad \tilde{M}=\lambda _{2}S\,, \qquad
\tilde{U}_{1}=\lambda _{2}T_{1}\,, \qquad \tilde{U}_{2}=\lambda _{2}T_{2}\,, \notag \\
\tilde{Z}&=&\lambda _{4}J\,,\qquad \tilde{Z}_{a}=\lambda _{4}G_{a}\,,\, \qquad \, \tilde{T}=\lambda _{4}S\,, \qquad
\tilde{B}_{1}=\lambda _{4}T_{1}\,, \qquad \tilde{B}_{2}=\lambda _{4}T_{2}\,, \notag \\
\tilde{N}&=&\lambda _{6}J\,, \qquad \tilde{N}_{a}=\lambda _{6}G_{a}\,,\qquad \ \tilde{V}=\lambda _{6}S\,, \qquad
\tilde{C}_{1}=\lambda _{6}T_{1}\,, \qquad \, \tilde{C}_{2}=\lambda _{6}T_{2}\,, \notag \\ 
\tilde{Q}_{\alpha }^{+}&=& \lambda _{1}Q_{\alpha }^{+}\,, \, \, \ \ \tilde{Q}_{\alpha }^{-}=\lambda _{1}Q_{\alpha }^{-}\,, \quad \ \ \tilde{\Sigma}
_{\alpha }^{+}=\lambda _{3}Q_{\alpha }^{+}\,,\quad \, \tilde{\Sigma}%
_{\alpha }^{-}=\lambda _{3}Q_{\alpha }^{-}\,,\quad \ 
 \tilde{R}_{\alpha
}=\lambda _{1}R_{\alpha }\,, \notag \\
 \tilde{W}_{\alpha
}&=&\lambda _{3}R_{\alpha }\,, \quad \, \tilde{\Xi}
_{\alpha }^{+}=\lambda _{5}Q_{\alpha }^{+}\,, \qquad \tilde{\Xi}
_{\alpha }^{-}=\lambda _{5}Q_{\alpha }^{-}\,, \quad   \tilde{K}_{\alpha
}=\lambda _{5}R_{\alpha }\,.
\end{eqnarray}
One can show that, using the multiplication law (\ref{SE6}) and the original
(anti-)commutation relations of the Nappi-Witten superalgebra (\ref{sNW}),
the expanded generators satisfy the (anti-)commutation relations \eqref{sEB}, \eqref{sMEB3} along with
\begin{eqnarray}
\left[ \tilde{J},\tilde{N}_{a}\right] &=&\epsilon _{ab}\tilde{N}_{b}\,,\text{
\ \ \ \ \ \ \ \ } \ \ \ \ \ \ \ \left[ \tilde{P}_{a},\tilde{Z}_{b}\right] =-\epsilon _{ab}%
\tilde{V}\,,\text{\ \ \ \ \ \ }\, \ \ \ \ \ \ \ \ \left[ \tilde{H},\tilde{Z}_{a}\right]
=\epsilon _{ab}\tilde{N}_{b}\,,\text{ \ \ }  \notag \\
\left[ \tilde{Z},\tilde{P}_{a}\right] &=&\epsilon _{ab}\tilde{N}%
_{b}\,,\qquad  \ \ \ \ \ \ \ \ \ \ \left[ \tilde{G}_{a},\tilde{N}_{b}\right] =-\epsilon
_{ab}\tilde{V}\,\,,\text{ \ \ \ \ } \ \ \ \ \ \ \ \ \left[ \tilde{N},\tilde{G}_{a}\right]
=\epsilon _{ab}\tilde{N}_{b}\,, \notag \\
 \left[ \tilde{J},\tilde{\Xi}_{\alpha }^{\pm }\right] &=&-
\frac{1}{2}\left( \gamma _{0}\right) _{\alpha }^{\text{ }\beta }\tilde{\Xi}_{\beta }^{\pm }\,,\qquad \left[ \tilde{H},\tilde{\Sigma}_{\alpha }^{\pm}\right] =-\frac{1}{2}
\left( \gamma _{0}\right) _{\alpha }^{\text{ }\beta }\tilde{\Xi}_{\beta }^{\pm }\,,\qquad  \left[ \tilde{Z},\tilde{Q}_{\alpha }^{\pm }\right] =\frac{1%
}{2}\left( \gamma _{0}\right) _{\alpha }^{\text{ }\beta }\tilde{\Xi}_{\beta }^{\pm }\,,  \notag \\
\left[ \tilde{J},\tilde{K}_{\alpha }\right]  &=&-\frac{1}{2}\left(
\gamma _{0}\right) _{\alpha }^{\text{ }\beta }\tilde{K}_{\beta }\,,\qquad \left[ \tilde{H},\tilde{W}_{\alpha }\right] =-\frac{1}{2%
}\left( \gamma _{0}\right) _{\alpha }^{\text{ }\beta }\tilde{K}_{\beta
}\,,\qquad \left[ \tilde{Z},\tilde{R}_{\alpha }\right]  =\frac{1}{2}%
\left( \gamma _{0}\right) _{\alpha }^{\text{ }\beta }\tilde{K}_{\beta}\,, \notag \\
\left[ \tilde{G}_{a},\tilde{\Xi}_{\alpha }^{+}\right]  &=&-\frac{1}{2}\left(
\gamma _{a}\right) _{\alpha }^{\text{ }\beta }\tilde{\Xi}_{\beta }^{-}\,,\qquad \left[ \tilde{P}_{a},\tilde{\Sigma}_{\alpha }^{+}\right] =-\frac{1}{2%
}\left( \gamma _{a}\right) _{\alpha }^{\text{ }\beta }\tilde{\Xi}_{\beta
}^{-}\,,\qquad \left[ \tilde{Z}_{a},\tilde{Q}_{\alpha }^{+}\right]  =\frac{1}{2}%
\left( \gamma _{a}\right) _{\alpha }^{\text{ }\beta }\tilde{\Xi}_{\beta}^{-}\,, \notag \\
\left[ \tilde{G}_{a},\tilde{\Xi}_{\alpha }^{-}\right]  &=&-\frac{1}{2}\left(
\gamma _{a}\right) _{\alpha }^{\text{ }\beta }\tilde{K}_{\beta }\,,\qquad \left[ \tilde{P}_{a},\tilde{\Sigma}_{\alpha }^{-}\right] =-\frac{1}{2%
}\left( \gamma _{a}\right) _{\alpha }^{\text{ }\beta }\tilde{K}_{\beta
}\,,\qquad \left[ \tilde{Z}_{a},\tilde{Q}_{\alpha }^{-}\right]  =\frac{1}{2}%
\left( \gamma _{a}\right) _{\alpha }^{\text{ }\beta }\tilde{K}_{\beta}\,, \notag \\
\left[ \tilde{S},\tilde{\Xi}_{\alpha }^{+}\right]  &=&-\frac{1}{2}\left(
\gamma _{0}\right) _{\alpha }^{\text{ }\beta }\tilde{K}_{\beta }\,,\qquad \left[ \tilde{M},\tilde{\Sigma}_{\alpha }^{+}\right] =-\frac{1}{2%
}\left( \gamma _{0}\right) _{\alpha }^{\text{ }\beta }\tilde{K}_{\beta
}\,,\qquad \left[ \tilde{T},\tilde{Q}_{\alpha }^{+}\right]  =\frac{1}{2}%
\left( \gamma _{0}\right) _{\alpha }^{\text{ }\beta }\tilde{K}_{\beta}\,, \notag \\
\left[ \tilde{Y}_{1},\tilde{\Xi}_{\alpha }^{\pm}\right]  &=&-\frac{1}{2}\left(
\gamma _{0}\right) _{\alpha \beta} \tilde{\Xi}_{\beta }^{\pm}\,,\qquad \left[ \tilde{U}_{1},\tilde{\Sigma}_{\alpha }^{\pm}\right] =-\frac{1}{2%
}\left( \gamma _{0}\right) _{\alpha \beta} \tilde{\Xi}_{\beta
}^{\pm}\,,\qquad \left[ \tilde{B}_{1},\tilde{Q}_{\alpha }^{\pm}\right]  =\frac{1}{2}%
\left( \gamma _{0}\right) _{\alpha \beta }\tilde{\Xi}_{\beta}^{\pm}\,, \notag \\
\left[ \tilde{Y}_{2},\tilde{\Xi}_{\alpha }^{+}\right]  &=&-\frac{1}{2}\left(
\gamma _{0}\right) _{\alpha \beta} \tilde{K}_{\beta }\,,\qquad \left[ \tilde{U}_{2},\tilde{\Sigma}_{\alpha }^{+}\right] =-\frac{1}{2%
}\left( \gamma _{0}\right) _{\alpha \beta }\tilde{K}_{\beta
}\,,\qquad \left[ \tilde{B}_{2},\tilde{Q}_{\alpha }^{+}\right]  =\frac{1}{2}%
\left( \gamma _{0}\right) _{\alpha \beta }\tilde{K}_{\beta}\,, \notag \\
\left[ \tilde{Y}_{1},\tilde{K}_{\alpha }\right]  &=&-\frac{1}{2}\left( \gamma _{0}\right) _{\alpha \beta }\tilde{K}_{\beta }\,,\qquad \left[ \tilde{U}_{1},\tilde{W}_{\alpha }\right] =-\frac{1}{2%
}\left( \gamma _{0}\right) _{\alpha \beta }\tilde{K}_{\beta
}\,,\qquad \left[ \tilde{B}_{1},\tilde{R}_{\alpha }\right]  =\frac{1}{2}%
\left( \gamma _{0}\right) _{\alpha \beta }\tilde{K}_{\beta}\,, \notag \\
\left\{ \tilde{Q}_{\alpha }^{-},\tilde{\Xi}_{\beta }^{-}\right\}
&=&\left\{ \tilde{\Sigma}_{\alpha }^{-},\tilde{\Sigma}_{\beta }^{-}\right\}=-\left( \gamma ^{0}C\right) _{\alpha \beta }\tilde{V} {+} \left( \gamma
^{0}C\right) _{\alpha \beta }\tilde{C}_{2}\,,  \notag \\
\left\{ \tilde{Q}_{\alpha }^{\pm },\tilde{\Xi}_{\beta }^{\mp }\right\}
&=&\left\{ \tilde{\Sigma}_{\alpha }^{+},\tilde{\Sigma}_{\beta }^{-}\right\}=-\left( \gamma ^{a}C\right) _{\alpha \beta }\tilde{N}_{a}\,,\notag
 \\
\left\{ \tilde{Q}_{\alpha }^{+},\tilde{\Xi}_{\beta }^{+}\right\}
&=&\left\{ \tilde{\Sigma}_{\alpha }^{+},\tilde{\Sigma}_{\beta }^{+}\right\}=-\left( \gamma ^{0}C\right) _{\alpha \beta }\tilde{N}-\left( \gamma
^{0}C\right) _{\alpha \beta }\tilde{C}_{1}\,,  \notag \\
\left\{ \tilde{Q}_{\alpha }^{+},\tilde{K}_{\beta }\right\}  &=&\left\{ \tilde{\Sigma}_{\alpha }^{+},\tilde{W}_{\beta }\right\}=\left\{ \tilde{\Xi}_{\alpha }^{+},\tilde{R}_{\beta }\right\}=-\left(
\gamma ^{0}C\right) _{\alpha \beta }\tilde{V}-\left( \gamma ^{0}C\right)
_{\alpha \beta }\tilde{C}_{2}\,.\label{GMEB2}
\end{eqnarray}
This superalgebra corresponds to a supersymmetric extension of the GMEB algebra introduced in \cite{Concha:2020sjt}. Note that this is a new NR superalgebra, unlike the previous ones which had already been presented in the literature in previous works. The GMEB superalgebra is characterized by the presence of three extra fermionic generators, which are1 $\tilde{\Xi}_{\alpha}^{\pm}$ and $\tilde{K}_{\alpha}$, with respect to the MEB superalgebra. $\tilde{C}_1$ and $\tilde{C}_2$ play the role of central charges, while $\tilde{B}_1$ and $\tilde{B}_2$ act non-trivially on the fermionic generators $\tilde{Q}_{\alpha}^{\pm}$ and $\tilde{R}_{\alpha}$.
The non-vanishing components of an invariant supertrace are given by \eqref{invt1}, \eqref{invMB}, and 
\begin{eqnarray}
\left\langle \tilde{G}_{a}\tilde{N}_{b}\right\rangle &=&\left\langle \tilde{P
}_{a}\tilde{Z}_{b}\right\rangle \ =\ \alpha_{3}\delta _{ab}\,,  \notag \\
\left\langle \tilde{J}\tilde{V}\right\rangle &=&\left\langle \tilde{H}\tilde{
T}\right\rangle \ = \ \left\langle \tilde{M}\tilde{Z}\right\rangle \ =\ \left\langle
\tilde{S}\tilde{N}\right\rangle \ = \ -\alpha_{3}\,,  \notag\\
\left\langle \tilde{Y}_{1}\tilde{C}_{2}\right\rangle &=&\left\langle \tilde{U}_{1}\tilde{B}_{2}\right\rangle
\,=\,\left\langle \tilde{B}_{1}\tilde{U}_{2}\right\rangle \ = \ \left\langle \tilde{C}_{1}\tilde{Y}_{2}\right\rangle \ =\ \alpha_3 \,,\notag \\
\left\langle \tilde{Q}_{\alpha }^{-}\tilde{\Xi}_{\beta }^{-}\right\rangle
&=&\left\langle \tilde{\Sigma}%
_{\alpha }^{+}\tilde{W}_{\beta }\right\rangle \,=\,\left\langle \tilde{\Xi}_{\alpha }^{+}\tilde{R}_{\beta }\right\rangle\ =\ \left\langle \tilde{Q}%
_{\alpha }^{+}\tilde{K}_{\beta }\right\rangle \ =\ \left\langle \tilde{\Sigma}_{\alpha }^{-}\tilde{\Sigma}_{\beta }^{-}\right\rangle \ =\ 2\alpha_3 C_{\alpha \beta } \,.\label{invGMEB}
\end{eqnarray}
 The NR one-form gauge connection for the GMEB superalgebra reads
\begin{eqnarray}
A &=&\tau \tilde{H}+e^{a}\tilde{P}_{a}+\omega \tilde{J}+\omega ^{a}
\tilde{G}_{a}+k\tilde{Z}+k^{a}\tilde{Z}_{a}+f\tilde{N} +f^{a}\tilde{N}_{a}+m\tilde{M}+s\tilde{S} \notag \\
&&+t\tilde{T}+v\tilde{V}+y_{1}\tilde{Y}_{1}+y_{2}\tilde{Y}_{2}+b_1\tilde{B}_{1}+b_{2}\tilde{B}
_{2}+u_{1}\tilde{U}_{1}+u_{2}\tilde{U}_{2}+c_{1}\tilde{C}_{1}+c_{2}\tilde{C}_{2} \notag \\
&&+{\psi }^{+}\tilde{Q}^{+}+{\psi }^{-}\tilde{Q}^{-}+{\xi }^{+}\tilde{\Sigma}
^{+}+{\xi }^{-}\tilde{\Sigma}^{-}+\zeta^{+}\tilde{\Xi}^{+}+\zeta^{-}\tilde{\Xi}^{-}+{\rho }\tilde{R}+{\chi }\tilde{W}+\kappa\tilde{K}\,.
\label{oneform}
\end{eqnarray}%
The corresponding NR CS action can be obtained by inserting the above gauge connection and the invariant supertrace given by \eqref{invt1}, \eqref{invMB}, and \eqref{invGMEB} into the general expression for the CS action in three
spacetime dimensions \eqref{CS}. The aforesaid action reads, up to boundary terms, as follows:
\begin{equation}\label{CSGMEB}
    I_{\text{GMEB}}=\alpha_{0}I_{0}+\alpha_{1}I_{1}+\alpha_{2}I_{2}+\alpha_{3}I_{3}\,,
\end{equation}
where $I_0$ and $I_1$ are given in \eqref{I0I1}, $I_2$ is given in \eqref{MEBI2}, while the term along $\alpha_3$ reads
\begin{eqnarray}
I_{3} &=&\int \left.\omega _{a}R^{a}\left( f^{b}\right)
+f_{a}R^{a}\left( \omega ^{b}\right) +e_{a}R^{a}\left( k^{b}\right)
+k_{a}R^{a}\left( e^{b}\right) -2sR\left( f\right) \right.  \notag \\
&&-\left. 2vR\left( \omega \right) -2mR\left( k\right) -2tR\left( \tau
\right)+2y_{1}dc_{2}+2u_{1}db_{2}+2y_{2}dc_{1}+2b_{1}du_{2} \right.  \notag \\
&&+\left. 2\bar{\psi}^{-}\nabla \zeta ^{-}+2
\bar{\zeta}^{-}\nabla \psi ^{-}+2
\bar{\xi}^{-}\nabla \xi ^{-}+2\bar{\psi}^{+}\nabla \kappa +2\bar{\zeta}^{+}\nabla
\rho+2\bar{\xi}^{+}\nabla \chi \right.\notag \\
&&+\left.2\bar{\rho}\nabla \zeta^{+}+2\bar{\chi}\nabla \xi^{+}+2\bar{\kappa}\nabla \psi^{+}\,. \right.\label{CSGMEB2}
\end{eqnarray}
In the above action, the expressions for $R(\omega)$ and $R^{a}(\omega^{b})$ were defined in \eqref{RR}, while those for $R(\tau)$, $R(k)$, $R^{a}(e^{b})$, and $R^{a}(k^{b})$ are given in \eqref{curvi21}. Furthermore,
\begin{eqnarray}
R\left( f\right) &=&df\,,\notag \\
R^{a}\left( f^{b}\right) &=&df^{a}+\epsilon ^{ac}\omega f_{c}+\epsilon
^{ac}\tau k_{c}+\epsilon ^{ac}ke_{c}+\epsilon ^{ac}f\omega _{c}\,, 
\end{eqnarray}%
and the covariant
derivatives of the spinor $1$-form fields appearing in $I_{3}$ are given by
\begin{eqnarray}
\nabla \zeta ^{+} &=&d\zeta ^{+}+\frac{1}{2}\omega \gamma _{0}\zeta ^{+}+\frac{1}{2%
}\tau \gamma _{0}\xi ^{+}+\frac{1}{2%
}k\gamma _{0}\psi ^{+}-\frac{1}{2}y_{1}\gamma _{0}\zeta ^{+}-\frac{1}{2}%
u_{1}\gamma _{0}\xi ^{+}-\frac{1}{2}%
b_{1}\gamma _{0}\psi ^{+}\,,  \notag \\
\nabla \zeta ^{-} &=&d\zeta ^{-}+\frac{1}{2}\omega \gamma _{0}\zeta ^{-}+\frac{1}{2%
}\tau \gamma _{0}\xi ^{-}+\frac{1}{2%
}k \gamma _{0}\psi ^{-}+\frac{1}{2}e^{a}\gamma _{a}\xi ^{+}+\frac{1}{2}%
\omega ^{a}\gamma _{a}\zeta ^{+}+\frac{1}{2}%
k^{a}\gamma _{a}\psi ^{+}\notag \\
&&+\frac{1}{2}y_{1}\gamma _{0}\zeta ^{-}+\frac{1}{2%
}u_{1}\gamma _{0}\xi ^{-}+\frac{1}{2%
}b_{1}\gamma _{0}\psi ^{-}\,,
\notag \\
\nabla \kappa &=&d\kappa +\frac{1}{2}\omega \gamma _{0}\kappa+\frac{1}{2}\tau
\gamma _{0}\chi+\frac{1}{2}k\gamma _{0}\rho+\frac{1}{2}\omega
^{a}\gamma _{a}\zeta ^{-}+\frac{1}{2}e^{a}\gamma _{a}\xi ^{-}+\frac{1}{2}k^{a}\gamma _{a}\psi ^{-} +\frac{1}{2}s\gamma _{0}\zeta ^{+}  \notag \\
&&+\frac{1}{2}m\gamma _{0}\xi
^{+}+\frac{1}{2}t\gamma _{0}\psi
^{+}-\frac{1}{2}y_{2}\gamma _{0}\zeta ^{+}-\frac{1}{2}u_{2}\gamma _{0}\xi ^{+}-\frac{1}{2}b_{2}\gamma _{0}\psi ^{+}-\frac{1}{2}y_{1}\gamma _{0}\kappa -\frac{1}{2}u_{1}\gamma _{0}\chi \notag \\
&&-\frac{1}{2}b_{1}\gamma _{0}\rho\,, \label{fermcurvGMEB}
\end{eqnarray}
along with \eqref{fermcurvEB} and \eqref{fermcurv}.
From (\ref{CSGMEB}), we see that the CS action contains four independent sectors. The new gauge fields $f_{a}$, $f$, $v$, $\zeta^{+}$, $\zeta^{-}$, and $\kappa$ appear explicitly in the
last term proportional to $\alpha_{3}$, which corresponds to the CS
action for the new NR generalized Maxwell superalgebra. Note that the GMEB superalgebra allows
to include a cosmological constant term along $\alpha_{3}$ different from the one appearing in the case of the extended NR supergravity presented in \cite{Concha:2020tqx}. One can see that the field equations imply the vanishing of the curvature two forms given by \eqref{fermcurvEB}, \eqref{curvi2}, \eqref{fermcurv}, \eqref{curvi3}, \eqref{fermcurvGMEB}, and
\begin{eqnarray}
F\left(f\right)&=&R\left(f\right)+\bar{\psi}^{+}\gamma^{0}\zeta^{+}+\frac{1}{2}\bar{\xi}^{+}\gamma^{0}\xi^{+}\,, \notag \\
F^{a}\left( f^{b}\right) &=&R^{a}\left(f^{b}\right)+\bar{\psi}^{+}\gamma^{a} \zeta^{-}+\bar{\psi}^{-}\gamma^{a} \zeta^{+}+\bar{\xi}^{+}\gamma^{a} \xi^{-}\,, \notag \\
F\left(v\right)&=&dv+\epsilon^{ac}\omega_a k_c+\epsilon^{ac}e_a k_c+\bar{\psi}^{-}\gamma^{0} \zeta^{-}+\frac{1}{2}\bar{\xi}^{-}\gamma^{0} \xi^{-}+\bar{\psi}^{+}\gamma^{0} \kappa+\bar{\xi}^{+}\gamma^{0} \chi+\bar{\zeta}^{+}\gamma^{0} \rho\,, \notag\\
F\left(c_{1}\right)&=&dc_{1}+\bar{\psi}^{+}\gamma^{0}\zeta^{+}+\frac{1}{2}\bar{\xi}^{+}\gamma^{0}\xi^{+}\,, \notag \\
F\left(c_{2}\right)&=&dc_{2}-\bar{\psi}^{-}\gamma^{0} \zeta^{-}-\frac{1}{2}\bar{\xi}^{-}\gamma^{0} \xi^{-}+\bar{\psi}^{+}\gamma^{0} \kappa++\bar{\xi}^{+}\gamma^{0} \chi+\bar{\zeta}^{+}\gamma^{0} \rho\,, \label{curvi4}
\end{eqnarray}
which are the GMEB supercurvatures. 
A natural subsequent step would be to continue expanding the Nappi-Witten superalgebra with bigger semigroups. As we will show in the next section, all NR superalgebras coming from the semigroup expansion of \eqref{sNW}, when the semigroup under consideration is of the $S_E$-type, can be written in a very compact and general way in terms of the original (anti-)commutation relations \eqref{sNW} and the elements of the relevant semigroup. 


\subsection{Generalized extended Bargmann supergravity}

As it was shown in \cite{Concha:2020sjt}, the extended Bargmann, the MEB, and the GMEB algebras can be seen as particular cases of the so-called generalized extended Bargmann (GEB$^{\left(N\right)}$) algebra, which corresponds to the NR version of the $\mathfrak{B}_{N+2}$ algebra enlarged with $U_1$ generators.
In this section, we present a supersymmetric extension of the GEB$^{\left(N\right)}$ algebra which can be obtained by performing an $S$-expansion of the super Nappi-Witten algebra \eqref{sNW}. Similarly to the purely bosonic case, the new NR superalgebra contains the extended Bargmann, the MEB, and the GMEB superalgebras previously introduced as particular cases. 

Let $S_{E}^{\left(2N\right)}=\lbrace\lambda_0,\lambda_1,\lambda_2,\ldots,\lambda_{2N-1},\lambda_{2N},\lambda_{2N+1}\rbrace$ be the relevant semigroup whose elements satisfy
\begin{equation}
\lambda _{\mu}\lambda _{\nu} =\left\{
\begin{array}{lcl}
\lambda _{\mu+\nu}\,\,\,\, & \mathrm{if}\,\,\,\,\mu+\nu \leq 2N\,, &
\\
\lambda _{2N+1}\,\, & \mathrm{if}\,\,\,\,\mu+\nu > 2N\,, &
\end{array}
\right.  \label{SEN}
\end{equation}
where $\lambda_{2N+1}=0_S$ is the zero element of the semigroup such that $0_S\lambda_{\mu}=0_S$. Let $S_{E}^{\left(2N\right)}=S_0\cup S_1$ be a semigroup decomposition with
\begin{eqnarray}
S_0&=&\lbrace \lambda_{2i}, \ i=0,\ldots,N\rbrace \cup\lambda_{2N+1}\,,\notag\\
S_1&=&\lbrace \lambda_{2m-1}, \ m=1,\ldots,N\rbrace\cup\lambda_{2N+1}\,.
\end{eqnarray}
Such decomposition is said to be resonant since it behaves as the subspace decomposition of the Nappi-Witten superalgebra \eqref{subspace},
\begin{eqnarray}
S_0\cdot S_0&\subset& S_0\,, \qquad \qquad \qquad S_0\cdot S_1\subset S_1\,, \qquad \qquad \qquad S_1\cdot S_1\subset S_0\,. 
\end{eqnarray}
Then, after performing a resonant $S_{E}^{\left(2N\right)}$-expansion to the super Nappi-Witten algebra \eqref{sNW} and considering a $0_S$-reduction, we find a new NR superalgebra. The expanded NR generators are related to the super Nappi-Witten ones through the semigroup elements as follows:
\begin{eqnarray}
 \tilde{J}^{\left(i\right)}&=&\lambda_{2i}J\,, \qquad \qquad \qquad \qquad \tilde{Q}_{\alpha}^{+\left(m\right)}=\lambda_{2m-1}Q_{\alpha}^{+}\,,\notag \\
\tilde{G}_{a}^{\left(i\right)}&=&\lambda_{2i}G_{a}\,, \qquad \qquad \qquad \quad \  \tilde{Q}_{\alpha}^{-\left(m\right)}=\lambda_{2m-1}Q_{\alpha}^{-}\,,\notag \\ \tilde{S}^{\left(i\right)}&=&\lambda_{2i}S\,, \qquad \qquad \qquad \qquad \ \, \tilde{R}_{\alpha}^{\left(m\right)}=\lambda_{2m-1}R_{\alpha}\,, \notag \\
\tilde{T}_{1}^{\left(i\right)}&=&\lambda_{2i}T_{1}\qquad \qquad \qquad \qquad \ \ \ \, \tilde{T}_{2}^{\left(i\right)}=\lambda_{2i}T_{2}\,. \label{gen1}
\end{eqnarray}
One can prove that the expanded generators satisfy the following (anti-)commutation relations:
\begin{eqnarray}
\left[ \tilde{J}^{\left(i\right)},\tilde{G}_{a}^{\left(j\right)}\right] &=&\epsilon _{ab}\tilde{G}_{b}^{\left(i+ j\right)}\,,\qquad \qquad \qquad  \ \ \ \,  \left[
\tilde{G}_{a}^{\left(i\right)},\tilde{G}_{b}^{\left(j\right)}\right] =-\epsilon _{ab}\tilde{S}^{\left(i+ j\right)}\,,  \notag \\
\left[ \tilde{J}^{\left(i\right)},\tilde{Q}_{\alpha }^{\pm \left(m\right)}\right] &=&-\frac{1}{2}\left( \gamma _{0}\right)
_{\alpha }^{\,\ \beta }\tilde{Q}_{\beta }^{\pm \left(i+ m\right) }\,,\quad \ \ \ \ \ \left[
\tilde{J}^{\left(i\right)},\tilde{R}_{\alpha }^{\left(m\right)}\right] =-\frac{1}{2}\left( \gamma _{0}\right) _{\alpha }^{%
\text{ }\beta }\tilde{R}_{\beta }^{\left(i+ m\right)}\,,  \notag \\
\left[ \tilde{G}_{a}^{\left(i\right)},\tilde{Q}_{\alpha }^{+\left(m\right)}\right] &=&-\frac{1}{2}\left( \gamma _{a}\right)
_{\alpha }^{\,\ \beta }\tilde{Q}_{\beta }^{-\left(i+ m\right)}\,,\quad \ \ \, \left[ \tilde{G}_{a}^{\left(i\right)},\tilde{Q}_{\alpha
}^{-\left(m\right)}\right] =-\frac{1}{2}\left( \gamma _{a}\right) _{\alpha }^{\text{ }%
\beta }\tilde{R}_{\beta }^{\left( i+ m\right)}\,,  \notag \\
\left[ \tilde{S}^{\left(i\right)},\tilde{Q}_{\alpha }^{+\left(m\right)}\right] &=&-\frac{1}{2}\left( \gamma _{0}\right)
_{\alpha }^{\text{ }\beta }\tilde{R}_{\beta }^{\left(i+ m\right)}\,,\qquad   \ \ \, \left[
\tilde{T}_{1}^{\left(i\right)},\tilde{Q}_{\alpha }^{\pm\left(m\right)}\right] = \pm \frac{1}{2}\left( \gamma _{0}\right) _{\alpha
\beta }\tilde{Q}_{\beta }^{\pm\left(i+ m\right)}\,,  \notag \\
\left[ \tilde{T}_{2}^{\left(i\right)},\tilde{Q}_{\alpha }^{+\left(m\right)}%
\right] &=& \frac{1}{2}\left( \gamma _{0}\right) _{\alpha \beta }\tilde{R}_{\beta }^{\left(i+ m\right)} \,,\qquad \quad \ \ \ \left[ \tilde{T}_{1}^{\left(i \right)},\tilde{R}_{\alpha }^{\left(m\right)}\right] = \frac{1}{2}\left( \gamma _{0}\right)
_{\alpha \beta }\tilde{R}_{\beta }^{\left(i+ m\right)} \,, \notag \\
\left\{ \tilde{Q}_{\alpha }^{+\left(m\right)},\tilde{Q}_{\beta }^{-\left(n\right)}\right\} &=&-\left( \gamma
^{a}C\right) _{\alpha \beta }\tilde{G}_{a}^{\left(m+ n-1\right)}\,,  \notag \\
\left\{ \tilde{Q}_{\alpha }^{+\left(m\right)},\tilde{Q}_{\beta }^{+\left(n\right)}\right\} &=&-\left( \gamma
^{0}C\right) _{\alpha \beta }\tilde{J}^{\left( m+n-1 \right)}-\left( \gamma ^{0}C\right) _{\alpha \beta
}\tilde{T}_{1}^{\left( m+n-1\right)}\,,\text{ \qquad }  \notag \\
\left\{ \tilde{Q}_{\alpha }^{-\left(m\right)},\tilde{Q}_{\beta }^{-\left(n\right)}\right\} &=&-\left( \gamma
^{0}C\right) _{\alpha \beta }\tilde{S}^{\left(m+n-1\right)} + \left( \gamma ^{0}C\right) _{\alpha \beta
}\tilde{T}_{2}^{\left(m+n-1\right)}\,,  \notag \\
\left\{ \tilde{Q}_{\alpha }^{+\left(m\right)},\tilde{R}_{\beta }^{\left(n\right)}\right\} &=&-\left( \gamma ^{0}C\right)
_{\alpha \beta }\tilde{S}^{\left( m+n-1 \right)}-\left( \gamma ^{0}C\right) _{\alpha \beta }\tilde{T}_{2}^{\left(m+n-1\right)}\,. \label{GEB}
\end{eqnarray}
This is obtained by using the multiplication law of the $S_{E}^{\left(2N\right)}$ semigroup \eqref{SEN} and the super Nappi-Witten (anti-)commutation relations \eqref{sNW}.

The expanded superalgebra above generalizes the extended Bargmann superalgebra \cite{Bergshoeff:2016lwr} and corresponds to the supersymmetric extension of the GEB$^{\left(N\right)}$ algebra introduced in \cite{Concha:2020sjt}. The NR superalgebra \eqref{GEB} contains $2N+2$ bosonic generators in addition to the GEB$^{\left(N\right)}$ bosonic generators and $3N$ fermionic charges. The extra bosonic generators $\tilde{T}_{1}
^{\left(i\right)}$ and $\tilde{T}_{2}^{\left(i\right)}$ act non-trivially on the fermionic charges $\tilde{Q}_{\alpha}^{\left(m\right)}$ if $i+m\leq N$. On the other hand, both $\tilde{T}_{1}
^{\left(N\right)}$ and $\tilde{T}_{2}^{\left(N\right)}$ are central charges. One can see that the case $N=1$ reproduces the extended Bargmann superalgebra, while the $N=2$ and $N=3$ ones correspond to the MEB and GMEB superalgebras, respectively. At the bosonic level, as it was shown in \cite{Concha:2020sjt}, the so-called $\mathfrak{B}_{N+2}$ algebra \cite{Edelstein:2006se,Izaurieta:2009hz} is the relativistic counterpart of the GEB$
^{\left(N\right)}$ algebra. Then, one could deduce that the present generalization is the respective NR version of the supersymmetric extension of the $\mathfrak{B}_{N+2}$ algebra \cite{Concha:2014xfa}.

The gauge connection one-form $A$ for the GEB$^{\left(N\right)}$ superalgebra reads
\begin{equation}
    A=\omega^{\left(i\right)}\tilde{J}^{\left( i\right)}+\omega^{a \left(i\right)}\tilde{G}_{a}^{\left(i\right)}+s^{\left(i\right)}\tilde{S}^{\left(i\right)}+t_{1}^{\left(i\right)}\tilde{T}_{1}^{\left( i \right)}+t_{2}^{\left(i\right)}T_{2}^{\left(i\right)}+\bar{\psi}^{+\left(m\right)}\tilde{Q}^{+\left(m\right)}+\bar{\psi}^{-\left(m\right)}\tilde{Q}^{-\left(m\right)}+\bar{\rho}^{\left(m\right)}\tilde{R}
   ^{\left(m\right)}\,. \label{AGEB}
\end{equation}
The presence of the additional bosonic generators $\lbrace \tilde{T}_{1}^{\left(i\right)},\tilde{T}_{2}^{\left(i\right)}\rbrace$ is required to ensure a non-degenerate invariant supertrace allowing the construction of a proper NR CS supergravity action. In particular, the GEB$^{\left(N\right)}$ superalgebra admits the following non-degenerate invariant tensor:
\begin{eqnarray}
\left\langle \tilde{G}_{a}^{\left(i\right)}\tilde{G}_{b}^{\left(j\right)}\right\rangle &=&\alpha_{i+ j}\delta_{ab}\,,\notag \\
\left\langle \tilde{J}^{\left(i\right)}\tilde{S}^{\left(j\right)}\right\rangle &=&-\alpha_{i+ j}\,,\notag \\
\left\langle \tilde{T}_{1}^{\left(i\right)} \tilde{T}_{2}^{\left(j\right)}\right\rangle &=&\alpha_{i+ j}\,,\notag\\
\left\langle \tilde{Q}_{\alpha }^{-\left(m\right)}\tilde{Q}_{\beta }^{-\left(n\right)}\right\rangle &=&2
\alpha_{m+n-1 }C_{\alpha \beta }\,=\,\left\langle \tilde{Q}_{\alpha }^{+\left(m\right)}%
\tilde{R}_{\beta }^{\left(n\right)}\right\rangle \,, \label{invtGEB}
\end{eqnarray}
where $i,j=0,1,\ldots,N$; $m,n=1,2,\ldots,N$, and $i+j<N+1$, $i+m<N+1$, and $m+n<N+1$. Here, the non-vanishing components of an invariant tensor for the GEB$^{\left(N\right)}$ superalgebra are obtained from the super Nappi-Witten ones \eqref{NWinv} using the definitions of the $S$-expansion method \cite{Izaurieta:2006zz}. Then, a GEB$^{\left(N\right)}$ CS supergravity action can be constructed by inserting the gauge connection one-form \eqref{AGEB} and the non-vanishing components of the invariant tensor \eqref{invtGEB} in the general expression for the CS action \eqref{CS}, that yields
\begin{eqnarray}
 I_{\text{GEB}^{\left(N\right)}}&=& \alpha_{i}I_{i} \,\,=\, \alpha_0 I_0 + \alpha_1 I_1 + \ldots + \alpha_N I_N \,, \label{CSGEB}
\end{eqnarray}
with
\begin{eqnarray}
 I_{i}&=& \int \omega _{a}^{\left( j\right) }d\omega
^{a\left( k\right) }\delta _{j+ k}^{i}+\epsilon ^{ac}\omega _{a}^{\left(
j\right) }\omega ^{\left( k\right) }\omega _{c}^{\left( l\right) }\delta
_{j+ k+ l}^{i}-2s^{\left( j\right) }d\omega ^{\left( k\right) }\delta
_{j+ k}^{i}+2t_{1}^{\left(j\right)}dt_{2}^{\left(k\right)}\delta_{j+ k}^{i}  \notag \\
&&+2\bar{\psi}^{-\left(m\right)}\nabla\psi^{-\left(n\right)}\delta_{m+n-1}^{i}+2\bar{\psi}^{+\left(m\right)}\nabla\rho^{\left(n\right)}\delta_{m+n-1}^{i}+2\bar{\rho}^{\left(m\right)}\nabla\psi^{+\left(n\right)}\delta_{m+n-1}^{i}\,, \label{IGEB}
\end{eqnarray}
where the covariant derivatives of the spinor 1-forms for the GEB$^{\left(N\right)}$ superalgebra read
\begin{eqnarray}
 \nabla\psi^{+\left(m\right)}&=&d\psi^{+\left(m\right)}+\frac{1}{2}\omega^{\left(i\right)}\gamma_{0}\psi^{+\left(n\right)}\delta_{i+n}^{m}-\frac{1}{2}t_{1}^{\left(i\right)}\gamma_{0}\psi^{+\left(n\right)}\delta_{i+n}^{m} \,, \notag \\
 \nabla\psi^{-\left(m\right)}&=&d\psi^{-\left(m\right)}+\frac{1}{2}\omega^{\left(i\right)}\gamma_{0}\psi^{-\left(n\right)}\delta_{i+n}^{m}+\frac{1}{2}\omega^{a\left(i\right)}\gamma_{a}\psi^{+\left(n\right)}\delta_{i+ n}^{m}+\frac{1}{2}t_{1}^{\left(i\right)}\gamma_{0}\psi^{-\left(n\right)}\delta_{i+ n}^{m}\,, \notag \\
 \nabla\rho^{\left(m\right)}&=&d\rho^{\left(m\right)}+\frac{1}{2}\omega^{\left( i \right)}\gamma_{0}\rho^{\left(n\right)}\delta_{i+ n}^{m}+\frac{1}{2}\omega^{a\left(i\right)}\gamma_{a}\psi^{-\left(n\right)}\delta_{i+ n}^{m}+\frac{1}{2}s^{\left(i\right)}\gamma_{0}\psi^{+\left(n\right)}\delta_{i+ n}^{m} \notag \\
 &&-\frac{1}{2}t_{2}^{\left(i\right)}\gamma_{0}\psi^{+\left(n\right)}\delta_{i+ n}^{m}-\frac{1}{2}t_1^{\left(i\right)}\gamma_{0}\rho^{\left(n\right)}\delta_{i+ n}^{m}\,. \label{GEBcurv}
\end{eqnarray}
The new NR CS action is invariant under the GEB$^{\left(N\right)}$ superalgebra \eqref{GEB} and contains $N+1$ independent sectors proportional to the $\alpha_i$'s. Interestingly, one can see that the GEB$^{\left(i\right)}$ CS action, for $i<N$, appears as a particular subcase. Indeed, the $I_0$ and $I_1$ CS terms describe the most general extended Bargamnn supergravity action, where $I_0$ corresponds to a NR exotic term. The explicit extended Bargmann CS action \eqref{CSEB} is recovered by identifying the gauge field one-forms as
\begin{eqnarray}
 \omega^{\left(0\right)}&=&\omega\,,\qquad \qquad \omega_{a}^{\left(0\right)}=\omega_{a}\,,\qquad \quad \ \ \, s^{\left(0\right)}=s\,, \qquad \qquad  t_1^{\left(0\right)}=y_1\,,\qquad \qquad t_2^{\left(0\right)}=y_2\,, \notag \\
  \omega^{\left(1\right)}&=&\tau\,,\qquad \qquad \, \omega_{a}^{\left(1\right)}=e_{a}\,,\qquad \qquad s^{\left(1\right)}=m\,, \qquad \quad \ \, t_1^{\left(1\right)}=u_1\,, \qquad \qquad
    t_2^{\left(1\right)}=u_2\,,\notag \\
   \psi^{+\left(1\right)}&=&\psi^{+}\,, \qquad \, \ \ \psi^{-\left(1\right)}=\psi^{-}\,, \qquad \quad \ \, \rho^{\left(1\right)}=\rho\,. \label{ident1}
\end{eqnarray}
On the other hand, the $I_2$ term along with $I_1$ and $I_0$ describe the MEB CS supergravity action \cite{Concha:2019mxx}. In particular, the $I_2$ CS action coincides with the one obtained in \eqref{MEBI2} by identifying the gauge field one-forms as in \eqref{ident1} together with
\begin{eqnarray}
   \omega^{\left(2\right)}&=&k\,,\qquad \qquad \, \omega_{a}^{\left(2\right)}=k_{a}\,,\qquad \qquad s^{\left(2\right)}=t\,, \qquad \qquad t_1^{\left(2\right)}=b_1\,,\qquad \qquad \, t_2^{\left(2\right)}=b_2\,,\notag \\
   \psi^{+\left(2\right)}&=&\xi^{+}\,, \qquad \ \ \ \psi^{-\left(2\right)}=\xi^{-}\,, \qquad \quad \ \ \rho^{\left(2\right)}=\chi\,. \label{ident2}
\end{eqnarray}
In addition, one can see that the GMEB CS supergravity theory is described by $I_0,\,I_1,\,I_2$, and $I_3$ by considering the redefinitions of the gauge fields as in \eqref{ident1}, \eqref{ident2}, and 
\begin{eqnarray}
   \omega^{\left(3\right)}&=&f\,,\qquad \qquad \, \omega_{a}^{\left(3\right)}=f_{a}\,,\qquad \qquad s^{\left(3\right)}=u\,, \qquad \qquad t_1^{\left(3\right)}=c_1\,,\qquad \qquad \, t_2^{\left(3\right)}=c_2\,,\notag \\
   \psi^{+\left(3\right)}&=&\zeta^{+}\,, \qquad \ \ \ \psi^{-\left(3\right)}=\zeta^{-}\,, \qquad \quad \ \ \rho^{\left(3\right)}=\kappa\,. \label{ident3}
\end{eqnarray}
Novel generalizations of the extended Bargmann supergravity theory are obtained for $N>3$ and correspond to supersymmetric extensions of the GEB$^{(N)}$ gravity theory presented in \cite{Concha:2020sjt}. Let us note that the equations of motion for a specific GEB$^{\left(i\right)}$ superalgebra are given by the vanishing of the curvature two-forms associated with the respective superalgebra, which are given by \eqref{GEBcurv} along with
\begin{eqnarray}
F\left( \omega^{\left(i\right)} \right) &=&d\omega^{i}+\frac{1}{2}\bar{\psi}^{+\left(m\right)}\gamma^{0} \psi^{-\left(n\right)}\delta_{m+n-1}^{i} \,,  \notag \\
F^{a}\left( \omega ^{b\left(i\right)}\right) &=&d\omega^{a\left(i\right)}+\epsilon^{ac}\omega^{\left(j\right)}\omega_{c}^{\left(k\right)}\delta_{j+k}^{i}+\bar{\psi}^{+\left(m\right)}\gamma^{a} \psi^{-\left(n\right)}\delta_{m+n-1}^{i}\,, \notag\\
F\left(s^{\left(i\right)}\right)&=&ds^{\left(i\right)}+\frac{1}{2}\bar{\psi}^{-\left(m\right)}\gamma^{0} \psi^{-\left(n\right)}\delta_{m+n-1}^{i}+\bar{\psi}^{+\left(m\right)}\gamma^{0} \rho^{\left(n\right)}\delta_{m+n-1}^{i}\,, \notag\\
F\left(t_{1}^{\left(i\right)}\right)&=&dt_{1}^{\left(i\right)}+\frac{1}{2}\bar{\psi}^{+\left(m\right)}\gamma^{0}\psi^{+\left(n\right)}\delta_{m+n-1}^{i}\,, \notag \\
F\left(t_{2}^{\left(i\right)}\right)&=&dt_{2}^{\left(i\right)}-\frac{1}{2}\bar{\psi}^{-\left(m\right)}\gamma^{0} \psi^{-\left(n\right)}\delta_{m+n-1}^{i}+\bar{\psi}^{+\left(m\right)}\gamma^{0} \rho^{\left(n\right)}\delta_{m+n-1}^{i}\,. \label{curvin}
\end{eqnarray}
The physical implications of the additional bosonic and fermionic content remains as an interesting open issue that shall be considered in a future work.


\section{Generalized extended Newton-Hooke supergravity theory and semigroup expansion method}\label{gnh}

In this section, we apply the $S$-expansion method to the Nappi-Witten superalgebra \eqref{sNW} to find a different family of NR superalgebras. In particular, we shall see that the extended Newton-Hooke superalgebra \cite{Ozdemir:2019tby} and its generalizations can be obtained considering $S_{L}^{(1)}$ and $S_{\mathcal{M}}^{(N)}$ as the relevant semigroups, respectively. The semigroup choice is not arbitrary and comes from the expansion relation presented at the level of the asymptotic symmetry. Indeed, as it was shown in \cite{Caroca:2019dds}, the conformal superalgebra can be obtained by expanding the super Virasoro one using $S_{L}^{(1)}$ as the relevant semigroup, while generalizations of the superconformal symmetry are found using $S_{\mathcal{M}}$. It is interesting to notice that the same semigroup used to relate diverse infinite-dimensional superalgebras can be considered at the NR level. Furthermore, we will show that the generalized extended Newtoon-Hooke superalgebras, which we have denoted as GNH$
^{(N)}$, are related to the GEB$^{(N)}$ ones \eqref{GEB} through an IW contraction. The construction of a NR CS supergravity action based on the GNH$^{(N)}$ superalgebra is also presented.


\subsection{Extended Newton-Hooke supergravity}\label{enhs}

From the super Nappi-Witten algebra (\ref{sNW}), one can obtain a supersymmetric extension of the Newton-Hooke algebra (precisely, the extended Newton-Hooke superalgebra obtained in \cite{Concha:2020tqx}) through the $S$-expansion method considering $S_{L}^{\left( 1\right)
}=\lbrace \lambda _{0},\lambda _{1},\lambda _{2}\rbrace $ as the relevant
semigroup, whose elements satisfy the following multiplication law:
\begin{equation}
\begin{tabular}{l|lll}
$\lambda _{2}$ & $\lambda _{2}$ & $\lambda _{2}$ & $\lambda _{2}$ \\
$\lambda _{1}$ & $\lambda _{2}$ & $\lambda _{1}$ & $\lambda _{2}$ \\
$\lambda _{0}$ & $\lambda _{0}$ & $\lambda _{2}$ & $\lambda _{2}$ \\ \hline
& $\lambda _{0}$ & $\lambda _{1}$ & $\lambda _{2}$%
\end{tabular}
\label{ml2}
\end{equation}%
with $\lambda _{2}=0_{S}$ being the zero element of the semigroup. Let $S_{L}^{(1)}=S_0\cup S_1$ be a decomposition of the semigroup%
\begin{eqnarray}
S_{0} &=&\left\{ \lambda _{0},\lambda _{1},\lambda _{2}\right\} \,,  \notag
\\
S_{1} &=&\left\{ \lambda _{1},\lambda _{2}\right\} \,,
\end{eqnarray}%
which is said to be resonant \cite{Izaurieta:2006zz} since it satisfies the same structure as the subspaces of the super Nappi-Witten algebra \eqref{subspace},
\begin{eqnarray}
S_0\cdot S_0&\subset& S_0\,, \notag \\
S_0\cdot S_1&\subset& S_1\,, \notag \\
S_1\cdot S_1&\subset& S_0\,. 
\end{eqnarray}
An expanded superalgebra spanned by $\{L,L_a,N,X_1,X_2,\tilde{L},\tilde{L}_a,\tilde{N},\tilde{X}_1,\tilde{X}_2,\mathcal{Q}_{\alpha}^{+},\mathcal{Q}_{\alpha}^{-},\mathcal{R}_{\alpha} \}$ is obtained after considering a resonant $S_{L}^{(1)}$-expansion of the super Nappi-Witten algebra and performing a $0_{S}$-reduction. The expanded generators are related to the super Nappi-Witten ones through the semigroup elements as
\begin{eqnarray}
L&=&\lambda_1 J\,, \qquad \qquad L_a=\lambda_1 G_a\,, \qquad \qquad N=\lambda_1 S\,, \notag \\
\tilde{L}&=&\lambda_0 J\,, \qquad \qquad \tilde{L}_a=\lambda_0 G_a\,, \qquad \qquad \tilde{N}=\lambda_0 S\,, \notag \\
\mathcal{Q}_{\alpha}^{+}&=&\lambda_1 Q_{\alpha}^{+}\,, \qquad \quad \mathcal{Q}_{\alpha}^{-}=\lambda_1 Q_{\alpha}^{-}\,, \qquad \quad \, \mathcal{R}_{\alpha}=\lambda_1 R_{\alpha}\,, \notag \\
X_1&=&\lambda_1 T_1\,, \qquad  \quad \ \ X_2=\lambda_1 T_2\,, \notag \\
\tilde{X}_1&=&\lambda_0 T_1\,, \qquad \quad \ \ \tilde{X}_2=\lambda_0 T_2 \,.
\end{eqnarray}
Then, one can see that the expanded NR generators satisfy two copies of the Nappi-Witten algebra, one of which is augmented by supersymmetry,
\begin{eqnarray}
\left[ L,L_{a}\right]  &=&\epsilon _{ab}L_{b}\,,\qquad \qquad \quad \ \ \ \ \ \left[
L_{a},L_{b}\right] =-\epsilon _{ab}N\,,  \notag \\
\left[ \tilde{L},\tilde{L}_{a}\right]  &=&\epsilon _{ab}\tilde{L}%
_{b}\,,\qquad \qquad \quad \ \ \ \ \left[ \tilde{L}_{a},\tilde{L}_{b}\right] =-\epsilon
_{ab}\tilde{N}\,,  \notag \\
\left[ L,\mathcal{Q}_{\alpha }^{\pm }\right]  &=&-\frac{1}{2}\left( \gamma _{0}\right)
_{\alpha }^{\,\ \beta }\mathcal{Q}_{\beta }^{\pm } \,,\quad \ \ \ \ \ \, \ \left[ L,\mathcal{R}_{\alpha }\right] =-\frac{1}{2}\left( \gamma _{0}\right)
_{\alpha }^{\,\ \beta }\mathcal{R}_{\beta }\,,  \notag \\
\left[ L_{a},\mathcal{Q}_{\alpha }^{+}\right]  &=&-\frac{1}{2}\left( \gamma _{a}\right)
_{\alpha }^{\,\ \beta }\mathcal{Q}_{\beta }^{-}\,,\qquad \ \, \left[ L_{a},\mathcal{Q}_{\alpha }^{-}\right] =-\frac{1}{2}\left( \gamma _{a}\right)
_{\alpha }^{\,\ \beta }\mathcal{R}_{\beta }\,,  \notag \\
\left[ N,\mathcal{Q}_{\alpha }^{+}\right]  &=&-\frac{1}{2}\left( \gamma _{0}\right)
_{\alpha }^{\,\ \beta }\mathcal{R}_{\beta }\,,\quad \ \, \ \ \ \left[ X_1,\mathcal{Q}_{\alpha }^{\pm}\right]  =\pm\frac{1}{2}\left( \gamma _{0}\right) _{\alpha \beta }\mathcal{Q}_{\beta }^{\pm}\,, \notag \\
\left[ X_2,\mathcal{Q}_{\alpha }^{+}%
\right] &=& \frac{1}{2}\left( \gamma _{0}\right) _{\alpha \beta }\mathcal{R}_{\beta } \,,\qquad \quad \ \, \left[ X_1,\mathcal{R}_{\alpha }\right] = \frac{1}{2}\left( \gamma _{0}\right)
_{\alpha \beta }\mathcal{R}_{\beta } \,, \notag \\
\left\{ \mathcal{Q}_{\alpha }^{+},\mathcal{Q}_{\beta }^{-}\right\} &=&-\left(
\gamma ^{a}C\right) _{\alpha \beta }L_{a}\,,  \notag \\
\left\{ \mathcal{Q}_{\alpha }^{+},\mathcal{Q}_{\beta }^{+}\right\}  &=& - \left( \gamma
^{0} C \right) _{\alpha \beta }L - \left( \gamma^{0}C\right) _{\alpha \beta }X_1   \,, \notag \\
\left\{ \mathcal{Q}_{\alpha }^{-},\mathcal{Q}_{\beta}^{-}\right\} &=&- \left( \gamma ^{0}C \right) _{\alpha \beta }N + \left( \gamma ^{0}C\right) _{\alpha \beta }X_2 ,  \notag \\
\left\{ \mathcal{Q}_{\alpha }^{+},\mathcal{R}_{\beta }\right\}  &=&- \left( \gamma ^{0}C\right)
_{\alpha \beta }N - \left( \gamma ^{0} C\right) _{\alpha \beta }X_2 \,,  \label{2NH}
\end{eqnarray}
where we have used the multiplication law of the semigroup $S_{L}^{(1)}$ (\ref{ml2}) and
the (anti-)commutation relations of the Nappi-Witten superalgebra \eqref{sNW}. On the other hand, let us note that the purely bosonic copy of the Nappi-Witen algebra is endowed with two $\mathfrak{u}(1)$ generators, namely $\tilde{X}_1$ and $\tilde{X}_2$. Considering the definitions of \cite{Izaurieta:2006zz}, one can see that the superalgebra \eqref{2NH} admits the following non-vanishing components of the invariant tensor:
\begin{eqnarray}
\left\langle L_{a}L_{b}\right\rangle &=&\mu\delta_{ab}\,, \qquad \qquad 
\left\langle \tilde{L}_{a}\tilde{L}_{b}\right\rangle =\nu\delta_{ab}\,,\notag \\
\left\langle LN\right\rangle &=&-\mu\,, \qquad \qquad \ \ \
\left\langle \tilde{L}\tilde{N}\right\rangle =-\nu\,,\notag \\
\left\langle X_1 X_2\right\rangle &=&\mu\,, \qquad \qquad \ \ \
\left\langle \tilde{X}_1 \tilde{X}_2\right\rangle =\nu\,, \notag \\
\left\langle \mathcal{Q}_{\alpha }^{-}\mathcal{Q}_{\beta }^{-}\right\rangle &=&2\mu
C_{\alpha \beta }\,, \qquad \ \ \, \left\langle \mathcal{Q}_{\alpha }^{+}%
\mathcal{R}_{\beta }\right\rangle=2\mu c_{\alpha \beta} \,,\label{invt2nw}
\end{eqnarray}
where $\mu$ and $\nu$ are arbitrary independent constants related to the ones of the super Nappi-Witten and Nappi-Witten algebras, respectively. Interestingly, the superalgebra \eqref{2NH} can be related to a central extension of the extended Bargmann
superalgebra \eqref{sEB} after a suitable redefinition of the NR generators and after having performed a vanishing cosmological constant limit. Indeed, let us consider the following redefinition of the generators:
\begin{eqnarray}
\tilde{G}_{a}&=&L_a-\tilde{L}_a\,, \qquad \qquad \ \  \tilde{P}_a=\frac{1}{\ell}\left(L_a+\tilde{L}_a \right)\,, \qquad \qquad \ \  \tilde{Q}_{\alpha}^{+}=\sqrt{\frac{2}{\ell}} \mathcal{Q}_{\alpha}^{+}\,,\notag \\
\tilde{J}&=&L+\tilde{L}\,, \qquad \qquad \ \ \ \ \ \tilde{H}=\frac{1}{\ell}\left(L-\tilde{L} \right)\,, \qquad \qquad \ \ \ \ \ \tilde{Q}_{\alpha}^{-}=\sqrt{\frac{2}{\ell}} \mathcal{Q}_{\alpha}^{-}\,,\notag \\
\tilde{S}&=&N+\tilde{N}\,, \qquad \qquad \ \ \ \, \tilde{M}=\frac{1}{\ell}\left(N-\tilde{N} \right)\,, \qquad \qquad \ \ \ \ \tilde{R}_{\alpha}=\sqrt{\frac{2}{\ell}} \mathcal{R}_{\alpha}\,,\notag \\
\tilde{Y}_{1}&=&X_1-\tilde{X}_1\,, \qquad \qquad \ \,    \tilde{U}_1=\frac{1}{\ell}\left(X_1+\tilde{X}_1 \right)\,, \notag \\
\tilde{Y}_{2}&=&X_2-\tilde{X}_2\,, \qquad \qquad \ \, \tilde{U}_2=\frac{1}{\ell}\left(X_2+\tilde{X}_2 \right)\,, \label{redef1b}
\end{eqnarray}
where $\ell$ is a length parameter related to the cosmological constant through $\Lambda \propto \pm \frac{1}{\ell^2}$. Thus, the expanded superalgebra (\ref{2NH}) can be rewritten as%
\begin{eqnarray}
\left[\tilde{J},\tilde{G}_a \right]&=&\epsilon_{ab}\tilde{G}_b\,, \qquad \qquad \ \ \left[ \tilde{G}_a,\tilde{G}_b \right]=-\epsilon_{ab}\tilde{S}\,,\qquad \qquad \quad \ \left[\tilde{J},\tilde{P}_a\right]=\epsilon_{ab}\tilde{P}_b\,, \notag\\
\left[\tilde{H},\tilde{G}_a \right]&=&\epsilon_{ab}\tilde{P}_b\,, \qquad \qquad \ \ \ \left[ \tilde{G}_a,\tilde{P}_b \right]=-\epsilon_{ab}\tilde{M}\,,\qquad \quad \quad \ \ \left[\tilde{H},\tilde{P}_a\right]=\frac{1}{\ell^{2}}\epsilon_{ab}\tilde{G}_b\,, \notag\\
\left[\tilde{P}_a,\tilde{P}_b \right]&=&-\frac{1}{\ell^{2}}\epsilon_{ab}\tilde{S}\,, \quad \qquad \ \ \left[ \tilde{J}, \tilde{Q}^{\pm}_\alpha \right]=- \frac{1}{2} \left(\gamma_0\right)_\alpha^{\phantom{\alpha} \beta} \tilde{Q}^{\pm}_\beta \,, \quad \ \left[ \tilde{H}, \tilde{Q}^{\pm}_\alpha \right]=- \frac{1}{2 \ell} \left(\gamma_0\right)_\alpha^{\phantom{\alpha} \beta} \tilde{Q}^{\pm}_\beta \,, \notag\\
\left[ \tilde{J}, \tilde{R}_\alpha \right]&=&- \frac{1}{2} \left(\gamma_0\right)_\alpha^{\phantom{\alpha} \beta} \tilde{R}_\beta \,, \ \ \ \ \left[ \tilde{H}, \tilde{R}_\alpha \right]=- \frac{1}{2 \ell} \left(\gamma_0\right)_\alpha^{\phantom{\alpha} \beta} \tilde{R}_\beta \,, \ \ \, \left[ \tilde{G}_a, \tilde{Q}^{+}_\alpha \right]=- \frac{1}{2} \left(\gamma_a\right)_\alpha^{\phantom{\alpha} \beta} \tilde{Q}^{-}_\beta \,, \notag \\
\left[ \tilde{P}_a, \tilde{Q}^{+}_\alpha \right]&=&- \frac{1}{2 \ell} \left(\gamma_a\right)_\alpha^{\phantom{\alpha} \beta} \tilde{Q}^{-}_\beta \,, \ \left[ \tilde{G}_a, \tilde{Q}^{-}_\alpha \right]=- \frac{1}{2} \left(\gamma_a\right)_\alpha^{\phantom{\alpha} \beta} \tilde{R}_\beta \,, \ \ \ \ \left[ \tilde{P}_a, \tilde{Q}^{-}_\alpha \right]=- \frac{1}{2 \ell} \left(\gamma_a\right)_\alpha^{\phantom{\alpha} \beta} \tilde{R}_\beta \,, \notag \\
\left[ \tilde{S}, \tilde{Q}^{+}_\alpha \right]&=&- \frac{1}{2} \left(\gamma_0\right)_\alpha^{\phantom{\alpha} \beta} \tilde{R}_\beta \,, \quad \, \left[ \tilde{M}, \tilde{Q}^{+}_\alpha \right]=- \frac{1}{2 \ell} \left(\gamma_0\right)_\alpha^{\phantom{\alpha} \beta} \tilde{R}_\beta \,, \ \ \ \left[ \tilde{Y}_1, \tilde{Q}^{\pm}_\alpha \right]= \pm \frac{1}{2} \left(\gamma_0\right)_{\alpha \beta} \tilde{Q}^{\pm}_\beta \,, \notag \\
\left[ \tilde{U}_1, \tilde{Q}^{\pm}_\alpha \right]&=& \pm \frac{1}{2 \ell} \left(\gamma_0\right)_{\alpha \beta} \tilde{Q}^{\pm}_\beta \,, \ \ \left[ \tilde{Y}_2, \tilde{Q}^{+}_\alpha \right]= \frac{1}{2} \left(\gamma_0\right)_{\alpha \beta} \tilde{R}_\beta \,, \quad \ \ \ \, \left[ \tilde{U}_2, \tilde{Q}^{+}_\alpha \right]= \frac{1}{2 \ell} \left(\gamma_0\right)_{\alpha \beta} \tilde{R}_\beta \,, \notag \\
\left[ \tilde{Y}_1, \tilde{R}_\alpha \right]&=& \frac{1}{2} \left(\gamma_0\right)_{\alpha \beta} \tilde{R}_\beta \,, \quad \quad \left[ \tilde{U}_1, \tilde{R}_\alpha \right]= \frac{1}{2 \ell} \left(\gamma_0\right)_{\alpha \beta} \tilde{R}_\beta \,, \notag \\
\left\{ \tilde{Q}_{\alpha }^{+},\tilde{Q}_{\beta }^{-} \right\} &=&- \frac{1}{\ell} \left( \gamma ^{a}C\right)
_{\alpha \beta } \tilde{G}_a - \left( \gamma ^{a}C\right)
_{\alpha \beta } \tilde{P}_a \,, \notag \\
\left\{ \tilde{Q}_{\alpha }^{+},\tilde{Q}_{\beta }^{+}\right\} &=&- \frac{1}{\ell} \left( \gamma ^{0}C\right)
_{\alpha \beta } \tilde{J} -\left( \gamma ^{0}C\right) _{\alpha \beta }\tilde{H} - \frac{1}{\ell} \left( \gamma ^{0}C\right)_{\alpha \beta } \tilde{Y}_1  -\left( \gamma ^{0}C\right) _{\alpha \beta }\tilde{U}_1 \,, \notag \\
\left\{ \tilde{Q}_{\alpha }^{-},\tilde{Q}_{\beta }^{-} \right\} &=&- \frac{1}{\ell} \left( \gamma ^{0}C\right)
_{\alpha \beta } \tilde{S} - \left( \gamma ^{0}C\right)
_{\alpha \beta } \tilde{M} + \frac{1}{\ell} \left( \gamma ^{0}C\right) _{\alpha \beta } \tilde{Y}_{2} + \left( \gamma ^{0}C\right) _{\alpha \beta } \tilde{U}_{2} \,, \notag \\
\left\{ \tilde{Q}_{\alpha }^{+},\tilde{R}_{\beta }\right\} &=&- \frac{1}{\ell} \left( \gamma ^{0}C\right)
_{\alpha \beta } \tilde{S} - \left( \gamma ^{0}C\right)
_{\alpha \beta } \tilde{M} - \frac{1}{\ell} \left( \gamma ^{0}C\right) _{\alpha \beta } \tilde{Y}_{2} - \left( \gamma ^{0}C\right) _{\alpha \beta } \tilde{U}_{2} \,. \label{sENH}
\end{eqnarray}
This superalgebra corresponds to a supersymmetric extension of the extended Newton-Hooke algebra \cite{Aldrovandi:1998im,Gibbons:2003rv,Brugues:2006yd,Alvarez:2007fw,Papageorgiou:2010ud,Duval:2011mi,Hartong:2016yrf,Duval:2016tzi} and coincides with the one presented in \cite{Concha:2020tqx}. One can notice that it contains two additional bosonic generators, namely $\tilde{Y}_{1}$ and $\tilde{Y}_{2}$, with respect to the extended Newton-Hooke superalgebra introduced in \cite{Ozdemir:2019tby}. Such generators, along with $\tilde{U}_{1}$ and $\tilde{U}_{2}$, act non-trivially on the fermionic generators $\tilde{Q}_{\alpha}
^{\pm}$ and $\tilde{R}_{\alpha}$. Remarkably, in the flat limit $\ell \rightarrow
\infty $ we recover the centrally extended version of the extended Bargmann superalgebra \eqref{sEB} with central charges $\tilde{U}_1$ and $\tilde{U}_2$. 

The presence of the additional bosonic generators ensures to have a non-degenerate invariant tensor allowing us to construct the most general CS action invariant under the extended Newton-Hooke superalgebra. In particular, the invariant tensor can be obtained from the invariant tensor of the two copies of the Nappi-Witten algebra, one of which is supersymmetric. Indeed, considering the redefinition of the generators \eqref{redef1b} in the invariant tensor \eqref{invt2nw}, we find that the extended Newton-Hooke superalgebra admits the following non-vanishing components of the invariant tensor:
\begin{eqnarray}
\left\langle \tilde{G}_{a} \tilde{G}_{b} \right\rangle&=& \alpha_0 \delta_{\alpha\beta}\,, \qquad \qquad \ \left\langle \tilde{J} \tilde{S} \right\rangle = -\alpha_0\,, \qquad \qquad \left\langle \tilde{Y}_{1} \tilde{Y}_{2} \right\rangle = \alpha_0\,, \notag \\
\left\langle \tilde{G}_{a} \tilde{P}_{b} \right\rangle&=& \alpha_1 \delta_{\alpha\beta}\,, \qquad \qquad \left\langle \tilde{J} \tilde{M} \right\rangle = -\alpha_1\,, \qquad \qquad \left\langle \tilde{Y}_{1} \tilde{U}_{2} \right\rangle = \alpha_1\,,\notag \\
\left\langle \tilde{P}_{a} \tilde{P}_{b} \right\rangle&=& \frac{\alpha_0}{\ell^{2}} \delta_{\alpha\beta}\,, \qquad \qquad \left\langle \tilde{H} \tilde{S} \right\rangle = -\alpha_1\,, \qquad \qquad \left\langle \tilde{U}_{1} \tilde{Y}_{2} \right\rangle = \alpha_1\,,\notag \\
\left\langle \tilde{H} \tilde{M} \right\rangle&=& -\frac{\alpha_0}{\ell^{2}}\,, \qquad \qquad   \left\langle \tilde{U}_{1} \tilde{U}_{2} \right\rangle = \frac{\alpha_0}{\ell^{2}}\,, \label{invtNH1}
\end{eqnarray}
along with
\begin{eqnarray}
\left\langle \tilde{Q}_{\alpha}^{-} \tilde{Q}_{\beta}^{-} \right\rangle &=&2\left(\frac{\alpha_0}{\ell}+\alpha_1\right)C_{\alpha\beta}\,, \notag \\
\left\langle \tilde{Q}_{\alpha}^{+} \tilde{R}_{\beta} \right\rangle &=&2\left(\frac{\alpha_0}{\ell}+\alpha_1\right)C_{\alpha\beta}\,, \label{invtNH2}
\end{eqnarray}
where the Newton-Hooke parameters are related to the Nappi-Witten ones through
\begin{equation}
    \alpha_0=\mu+\nu\,, \qquad \qquad \alpha_1=\frac{1}{\ell}\left(\mu-\nu\right)\,.
\end{equation}
One can see that the invariant tensor \eqref{invt1} for the extended Bargmann superalgebra are recovered in the vanishing cosmological constant limit $\ell \rightarrow \infty$. 

Let us now consider the gauge connection one-form $A$ for the extended Newton-Hooke superalgebra \eqref{sENH}, that is
\begin{align}
A=\ &\omega \tilde{J}+\tau\tilde{H}+\omega^{a}\tilde{G}_{a}+e^{a}\tilde{P}_{a}+s\tilde{S}+m\tilde{M}+y_{1}\tilde{Y}_{1}+y_{2}\tilde{Y}_{2}+u_{1}\tilde{U}_{1}+u_{2}\tilde{U}_{2}\notag\\
&+\bar{\psi}^{+}\tilde{Q}^{+}+\bar{\psi}^{-}\tilde{Q}^{-}+\bar{\rho}\tilde{R}\,. \label{1fNH}
\end{align}
Then, the three-dimensional extended Newton-Hooke CS supergravity theory is obtained considering the gauge connection one-form \eqref{1fNH} and the non-vanishing components of the invariant tensor \eqref{invtNH1}-\eqref{invtNH2} and plugging them into the general expression for a CS action \eqref{CS}. By doing so, one gets
\begin{equation}
    I_{\text{NH}}=\alpha_{0}I_{0}+\alpha_{1}I_{1}\,,\label{CSNH}
\end{equation}
where
\begin{eqnarray}
I_{0}&=&\int\omega _{a}R^{a}(\omega ^{b})-2sR\left(
\omega \right)+\frac{1}{\ell^{2}}e_{a}R^{a}\left(e^{b}\right)-\frac{1}{\ell^{2}}2mR\left(\tau\right) +2y_{1}dy_{2}+\frac{2}{\ell^{2}}u_{1}du_{2} \notag\\
&&-\frac{2}{\ell}\bar{\psi}^{-}\nabla\psi^{-}-\frac{2}{\ell}\bar{\psi}^{+}\nabla\rho-\frac{2}{\ell}\bar{\rho}\nabla\psi^{+}\,,\label{CSNH0} \\
I_{1}&=&\int2e_{a}R^{a}(\omega ^{b})-2mR(\omega )-2\tau R(s)+\frac{1}{\ell^{2}}\epsilon_{ab}\tau e^{a}e^{b}+2y_{1}du_{2}+2u_{1}dy_{2}  \notag \\
&& -2\bar{\psi}^{-}\nabla \psi^{-}-2\bar{\psi}^{+}\nabla \rho- 2\bar{\rho}\nabla \psi ^{+}\,.\label{CSNH1}
\end{eqnarray}
Here, the curvature two-form $R(s)$ is given by
\begin{equation}
R(s)=ds +\frac{1}{2}\epsilon^{ac}\omega_a\omega_c\,,
\end{equation}
while $R(\omega)$ and $R^{a}(\omega^{b})$ are given by \eqref{RR}. Besides, the covariant derivatives of the spinor 1-form fields read
\begin{eqnarray}
\nabla \psi ^{+} &=&d\psi^{+}+\frac{1}{2}\omega \gamma _{0}\psi^{+}+\frac{1}{2\ell}\tau\gamma_0\psi^{+}-\frac{1
}{2}y_{1}\gamma _{0}\psi^{+}-\frac{1}{2\ell}u_{1}\gamma_{0}\psi^{+}\,,  \notag \\
\nabla \psi ^{-} &=&d\psi^{-}+\frac{1}{2}\omega \gamma _{0}\psi^{-}+\frac{1
}{2}\omega ^{a}\gamma _{a}\psi ^{+}+\frac{1}{2\ell}\tau\gamma_{0}\psi^{-}+\frac{1}{2\ell}e^{a}\gamma_{a}\psi^{+}+\frac{1}{2}y_{1}\gamma _{0}\psi^{-}+\frac{1}{2\ell}u_{1}\gamma_{0}\psi^{-}\,, \notag \\
\nabla \rho &=&d\rho +\frac{1}{2}\omega \gamma _{0}\rho +\frac{1}{2}\omega
^{a}\gamma _{a}\psi ^{-}+\frac{1}{2}s\gamma _{0}\psi^{+}+\frac{1}{2\ell}\tau\gamma_{0}\rho+\frac{1}{2\ell}e^{a}\gamma_{a}\psi^{-}+\frac{1}{2\ell}m\gamma_{0}\psi^{+}-\frac{1}{2}
y_{2}\gamma _{0}\psi ^{+}\,,\notag \\
&&-\frac{1}{2}y_{1}\gamma _{0}\rho-\frac{1}{2\ell}u_{2}\gamma_{0}\psi^{+}-\frac{1}{2\ell}u_1\gamma_{0}\rho\,.  \label{spinorcurv}
\end{eqnarray}
The three-dimensional CS extended Newton-Hooke supergravity action \eqref{CSNH} can be written in terms of two independent CS actions. The CS term $I_0$ corresponds to an exotic CS NR supergravity action \cite{Ozdemir:2019tby} and can be seen as the NR version of the exotic $\mathcal{N}=2$ CS AdS supergravity action \cite{Achucarro:1987vz,Howe:1995zm,Giacomini:2006dr},
\begin{eqnarray}
I_{\text{exotic}}^{\text{AdS}}&=&\int \omega_A d\omega^{A}+\frac{1}{3}\epsilon_{ABC}\omega^{A}\omega^{B}\omega^{C}+\frac{1}{\ell^{2}}e_{A}T^{A} +\mathsf{t}d\mathsf{t}+\frac{1}{\ell^{2}}\mathsf{u}d\mathsf{u} -\frac{2}{\ell}\bar{\psi}^{i}\nabla\psi^{i} \,,  \label{exoticAdS}
\end{eqnarray}
where $A=0,1,2$, $i=1,2$, and $\lbrace \mathsf{t}, \mathsf{u}\rbrace$ are $\mathfrak{so}(2)$ internal symmetry gauge fields. On the other hand, the CS term $I_1$ is the supersymmetric extension of the Newton-Hooke gravity action and resembles the extended Newton-Hooke supergravity action introduced in \cite{Ozdemir:2019tby} except for the presence of the additional bosonic gauge fields $y_1$, $y_2$, $u_1$, and $u_2$. At the bosonic level, the relativistic exotic term is related to the Pontryagin density, while the term along $\alpha_1$ corresponds to an Euler CS form \cite{Troncoso:1999pk}. The combination of both families allows to write the most general CS action not only for the AdS (or Poincaré) algebra but also for the Maxwell-like symmetries \cite{Concha:2014zsa}. Interestingly, the exotic Pontryagin CS term can be extended at the supersymmetric and NR level, corresponding to the $I_0$ CS action. As we shall see in the next section, the NR exotic terms can also be introduced into a generalized extended Newton-Hooke supergravity theory.

Let us note that the non-degeneracy of the invariant tensor \eqref{invtNH1}-\eqref{invtNH2} is related to the requirement that the CS supergravity action involves a kinematical term for each gauge field. Then, the equations of motion of the extended Newton-Hooke supergravity theory are given by the vanishing of the corresponding curvature two-forms, which are given by \eqref{spinorcurv} along with
\begin{eqnarray}
F\left( \omega \right) &=&R(\omega)+\frac{1}{2\ell}\bar{\psi}^{+}\gamma^{0} \psi^{+} \,,  \qquad \qquad \qquad \qquad \qquad \ \, F\left( \tau \right) =d\left( \tau \right)+\frac{1}{2}\bar{\psi}^{+}\gamma^{0} \psi^{-} \,,
 \notag\\
F^{a}\left( \omega ^{b}\right) &=&R^{a}(\omega^{b})+\frac{1}{\ell^{2}}\epsilon^{ac}\tau e_c+\frac{1}{2}\bar{\psi}^{+}\gamma^{0} \psi^{-}\,, \qquad \qquad \ \ \, F\left(y_{1}\right)=dy_{1}+\frac{1}{2\ell}\bar{\psi}^{+}\gamma^{0} \psi^{+} \,,\notag \\ F\left(s\right)&=&R(s)+\frac{1}{2\ell}\epsilon^{ac}e_a e_c+\frac{1}{2\ell}\bar{\psi}^{-}\gamma^{0} \psi^{-}+\frac{1}{\ell}\bar{\psi}^{+}\gamma^{0} \rho\,, \   F\left(y_{2}\right)=dy_{2}-\frac{1}{2\ell}\bar{\psi}^{-}\gamma^{0} \psi^{-}+\frac{1}{\ell}\bar{\psi}^{+}\gamma^{0} \rho\,,
  \notag \\
F^{a}\left( e ^{b}\right) &=&de^{a}+\epsilon^{ac}\omega e_c+\epsilon^{ac}\tau \omega_c+\bar{\psi}^{+}\gamma^{a} \psi^{-}\,, \qquad \qquad \, F\left(u_{2}\right)=du_{2}-\frac{1}{2}\bar{\psi}^{-}\gamma^{0} \psi^{-}+\bar{\psi}^{+}\gamma^{0} \rho \,,  \notag \\
F\left(m\right)&=&dm+\epsilon^{ac}\omega_a e_c+\frac{1}{2}\bar{\psi}^{-}\gamma^{0} \psi^{-}+\bar{\psi}^{+}\gamma^{0} \rho\,, \qquad  \quad F\left(u_{1}\right)=du_{1}+\frac{1}{2}\bar{\psi}^{+}\gamma^{0}\psi^{+} \,  
\,. \label{curvinh}
\end{eqnarray}
Here, one can prove that the vanishing cosmological constant limit $\ell\rightarrow\infty$ reproduces the equations of motion of the extended Bargmann supergravity theory. Thus,the extended Newton-Hooke supergravity theory reproduces, in the flat limit, the extended Bargmann supergravity theory \eqref{CSEB}. In particular, $I_0$ reduces to the exotic extended Bargamnn gravity action, while the $I_1$ term leads us to the usual extended Bargmann supergravity  \cite{Bergshoeff:2016lwr} endowed with extra bosonic gauge fields.


\subsection{Generalized extended Newton-Hooke supergravity}

A generalization of the extended Newton-Hooke superalgebra, denoted as GNH$^{(N)}$, can be obtained by performing an $S$-expansion of the super Nappi-Witten algebra \eqref{sNW}. To this end, let us first consider $S_{\mathcal{M}}^{(2N)}=\lbrace \lambda_0,\lambda_1,\lambda_2,\ldots,\lambda_{2N-1},\lambda_{2N} \rbrace$ as the relevant semigroup whose elements satisfy the following multiplication law:
\begin{equation}
\lambda _{\alpha }\lambda _{\beta }=\left\{
\begin{array}{lcl}
\lambda _{\alpha +\beta }\,\,\,\, & \mathrm{if}\,\,\,\,\alpha +\beta \leq 2N\,, &
\\
\lambda _{\alpha+\beta-2N}\,\, & \mathrm{if}\,\,\,\,\alpha +\beta > 2N\,, &
\end{array}
\right.  \label{SMN}
\end{equation}
with $N\geq2$. Let us note that, unlike the semigroups $S_{E}^{\left(2N\right)}$ and $S_{L}^{(1)}$, there is no zero element in the present semigroup. Let us now consider a semigroup decomposition $S_{\mathcal{M}}^{(2N)}=S_{0}\cup S_{1}$ where
\begin{eqnarray}
S_0&=&\lbrace \lambda_{2i}, \ i=0,\ldots,N\rbrace \,,\notag\\
S_1&=&\lbrace \lambda_{2m-1}, \ m=1,\ldots,N\rbrace\,.
\end{eqnarray}
Such decomposition is said to be resonant since it satisfies the same subspace decomposition of the Nappi-Witten superalgebra \eqref{subspace},
\begin{eqnarray}
S_0\cdot S_0&\subset& S_0\,, \notag \\
S_0\cdot S_1&\subset& S_1\,, \notag \\
S_1\cdot S_1&\subset& S_0\,. 
\end{eqnarray}
One can see that a novel family of NR superalgebras is obtained after performing a resonant $S_{\mathcal{M}}^{(2N)}$-expansion to the super Nappi-Witten algebra \eqref{sNW}. In particular, the expanded NR generators are related to the super Nappi-Witten ones through the semigroup elements as follows:
\begin{eqnarray}
 \tilde{J}^{(i)}&=&\lambda_{2i}J\,, \qquad \qquad \qquad \qquad \tilde{Q}_{\alpha}^{+(m)}=\lambda_{2m-1}Q_{\alpha}^{+}\,,\notag \\
\tilde{G}_{a}^{(i)}&=&\lambda_{2i}G_{a}\,, \qquad \qquad \qquad \quad \  \tilde{Q}_{\alpha}^{-(m)}=\lambda_{2m-1}Q_{\alpha}^{-}\,,\notag \\ \tilde{S}^{(i)}&=&\lambda_{2i}S\,, \qquad \qquad \qquad \qquad \ \, \tilde{R}_{\alpha}^{(m)}=\lambda_{2m-1}R_{\alpha}\,, \notag \\
\tilde{T}_{1}^{\left(i\right)}&=&\lambda_{2i}T_{1}\qquad \qquad \qquad \qquad \quad \,   \tilde{T}_{2}^{\left(i\right)}=\lambda_{2i}T_{2}\,. \label{gen2}
\end{eqnarray}
Then, considering the multiplication law of the $S_{\mathcal{M}}^{(2N)}$ semigroup \eqref{SMN} and the original commutation relations of the super Nappi-Witten algebra \eqref{sNW}, one can then show that the expanded NR superalgebra satisfy the following (anti-)commutation relations:
\begin{eqnarray}
\left[ \tilde{J}^{\left(i\right)},\tilde{G}_{a}^{\left(j\right)}\right] &=&\epsilon _{ab}\tilde{G}_{b}^{\left(i\ast j\right)}\,,\qquad \qquad \qquad  \ \ \  \,  \left[
\tilde{G}_{a}^{\left(i\right)},\tilde{G}_{b}^{\left(j\right)}\right] =-\epsilon _{ab}\tilde{S}^{\left(i\ast j\right)}\,,  \notag \\
\left[ \tilde{J}^{\left(i\right)},\tilde{Q}_{\alpha }^{\pm \left(m\right)}\right] &=&-\frac{1}{2}\left( \gamma _{0}\right)
_{\alpha }^{\,\ \beta }\tilde{Q}_{\beta }^{\pm \left(i\ast m\right) }\,,\quad \ \ \ \ \ \left[
\tilde{J}^{\left(i\right)},\tilde{R}_{\alpha }^{\left(m\right)}\right] =-\frac{1}{2}\left( \gamma _{0}\right) _{\alpha }^{%
\text{ }\beta }\tilde{R}_{\beta }^{\left(i\ast m\right)}\,,  \notag \\
\left[ \tilde{G}_{a}^{\left(i\right)},\tilde{Q}_{\alpha }^{+\left(m\right)}\right] &=&-\frac{1}{2}\left( \gamma _{a}\right)
_{\alpha }^{\,\ \beta }\tilde{Q}_{\beta }^{-\left(i\ast m\right)}\,,\quad \ \ \, \left[ \tilde{G}_{a}^{\left(i\right)},\tilde{Q}_{\alpha
}^{-\left(m\right)}\right] =-\frac{1}{2}\left( \gamma _{a}\right) _{\alpha }^{\text{ }%
\beta }\tilde{R}_{\beta }^{\left( i\ast m\right)}\,,  \notag \\
\left[ \tilde{S}^{\left(i\right)},\tilde{Q}_{\alpha }^{+\left(m\right)}\right] &=&-\frac{1}{2}\left( \gamma _{0}\right)
_{\alpha }^{\text{ }\beta }\tilde{R}_{\beta }^{\left(i\ast m\right)}\,,\qquad   \ \ \, \left[
\tilde{T}_{1}^{\left(i\right)},\tilde{Q}_{\alpha }^{\pm\left(m\right)}\right] = \pm \frac{1}{2}\left( \gamma _{0}\right) _{\alpha
\beta }\tilde{Q}_{\beta }^{\pm\left(i\ast m\right)}\,,  \notag \\
\left[ \tilde{T}_{2}^{\left(i\right)},\tilde{Q}_{\alpha }^{+\left(m\right)}%
\right] &=& \frac{1}{2}\left( \gamma _{0}\right) _{\alpha \beta }\tilde{R}_{\beta }^{\left(i\ast m\right)} \,,\qquad \quad \ \ \ \left[ \tilde{T}_{1}^{\left(i \right)},\tilde{R}_{\alpha }^{\left(m\right)}\right] = \frac{1}{2}\left( \gamma _{0}\right)
_{\alpha \beta }\tilde{R}_{\beta }^{\left(i\ast m\right)} \,, \notag \\
\left\{ \tilde{Q}_{\alpha }^{+\left(m\right)},\tilde{Q}_{\beta }^{-\left(n\right)}\right\} &=&-\left( \gamma
^{a}C\right) _{\alpha \beta }\tilde{G}_{a}^{\left(m\ast [n-1]\right)}\,,  \notag \\
\left\{ \tilde{Q}_{\alpha }^{+\left(m\right)},\tilde{Q}_{\beta }^{+\left(n\right)}\right\} &=&-\left( \gamma
^{0}C\right) _{\alpha \beta }\tilde{J}^{\left( m\ast [n-1] \right)}-\left( \gamma ^{0}C\right) _{\alpha \beta
}\tilde{T}_{1}^{\left( m\ast [n-1]\right)}\,,\text{ \qquad }  \notag \\
\left\{ \tilde{Q}_{\alpha }^{-\left(m\right)},\tilde{Q}_{\beta }^{-\left(n\right)}\right\} &=&-\left( \gamma
^{0}C\right) _{\alpha \beta }\tilde{S}^{\left(m\ast [n-1]\right)} + \left( \gamma ^{0}C\right) _{\alpha \beta
}\tilde{T}_{2}^{\left(m\ast [n-1]\right)}\,,  \notag \\
\left\{ \tilde{Q}_{\alpha }^{+\left(m\right)},\tilde{R}_{\beta }^{\left(n\right)}\right\} &=&-\left( \gamma ^{0}C\right)
_{\alpha \beta }\tilde{S}^{\left( m\ast [n-1] \right)}-\left( \gamma ^{0}C\right) _{\alpha \beta }\tilde{T}_{2}^{\left(m \ast [n-1]\right)}\,, \label{GNH}
\end{eqnarray}
where we have defined
\begin{equation}
    i \ast j=\left\{
    \begin{array}{lcl}
    i+j\,\,\,\, & \mathrm{if}\,\,\,\,i+j \leq N\,, &
    \\
    i+j-N\,\, & \mathrm{if}\,\,\,\,i+j > N\,. &
    \end{array}
    \right.
\end{equation}
The NR superalgebra \eqref{GNH} corresponds to a generalized extended Newton-Hooke superalgebra which we have denoted as GNH$^{(N)}$ superalgebra. The GNH$^{(N)}$ superalgebra is characterized by $3N$ fermionic generators and can be seen as the supersymmetric extension of the generalized Newton-Hooke algebra presented in \cite{Penafiel:2019czp}.  Furthermore, as a consequence of the $S$-expansion procedure, the GNH$^{(N)}$ superalgebra contains $2N$ additional bosonic generators with respect to the generalized Newton-Hooke ones. As we shall see, their presence ensures the non-degeneracy of the invariant tensor. The present superalgebra can be seen as the NR counterpart of the supersymmetric extension of the $\mathfrak{C}_{N+2}$ algebra \cite{Salgado:2014qqa, Concha:2016hbt}. However, the construction of a NR limit reproducing the GNH$
^{\left(N\right)}$ superalgebra shall not be explored here.

It is important to clarify that $N\geq 2$, while the usual extended Newton-Hooke superalgebra \eqref{sENH} does not appear as a particular subcase of GNH$^{\left(N\right)}$ since it has been obtained with a different semigroup, namely $S_{L}^{(1)}$. Nevertheless, since at the bosonic level the extended Newton-Hooke algebra belongs to the bosonic generalized Newton-Hooke family, we shall consider the extended Newton-Hooke superalgebra as the GNH$
^{\left(1\right)}$ superalgebra. The first case obtained with the present procedure corresponds to the GNH$^{(2)}$ superalgebra. In particular, the $N=2$ case reproduces the enlarged extended Bargmann (EEB) superalgebra recently introduced in \cite{Concha:2020tqx}. Remarkably, similarly to the extended Newton-Hooke superalgebra, the EEB superalgebra can be written as three copies of the Nappi-Witten algebra, two of which are augmented by supersymmetry \cite{Concha:2020tqx}. One could argue that the GNH$^{\left(N\right)}$ superalgebra could be written as $N+1$ copies of Nappi-Witten algebras, $N$ of which should be augmented by supersymmetry. Nevertheless, this does not hold true anymore for $N>2$. 

Let us note that the $2N$ extra bosonic generators $\tilde{T}_{1}^{\left(i\right)}$ and $\tilde{T}_{2}^{\left(i\right)}$ allow to have a non-degenerate invariant tensor which is essential to the proper construction of a NR CS supergravity action. Indeed, following the definitions of \cite{Izaurieta:2006zz}, it is possible to show that the GNH$^{\left(N\right)}$ superalgebra admits the following non-vanishing components of the invariant tensor:
\begin{eqnarray}
\left\langle \tilde{G}_{a}^{\left(i\right)}\tilde{G}_{b}^{\left(j\right)}\right\rangle &=&\alpha_{i\ast j}\delta_{ab}\,,\notag \\
\left\langle \tilde{J}^{\left(i\right)}\tilde{S}^{\left(j\right)}\right\rangle &=&-\alpha_{i\ast j}\,,\notag \\
\left\langle \tilde{T}_{1}^{\left(i\right)} \tilde{T}_{2}^{\left(j\right)}\right\rangle &=&\alpha_{i\ast j}\,,\notag\\
\left\langle \tilde{Q}_{\alpha }^{-\left(m\right)}\tilde{Q}_{\beta }^{-\left(n\right)}\right\rangle &=&2
\alpha_{m\ast (n-1) }C_{\alpha \beta }\,=\,\left\langle \tilde{Q}_{\alpha }^{+\left(m\right)}%
\tilde{R}_{\beta }^{\left(n\right)}\right\rangle \,, \label{invtGNH}
\end{eqnarray}
where the NR parameters appear as a consequence of the $S$-expansion procedure \cite{Izaurieta:2006zz}. On the other hand, the GNH$^{\left(N\right)}$ gauge connection one-form $A$ reads
\begin{equation}
    A=\omega^{\left(i\right)}\tilde{J}^{\left( i\right)}+\omega^{a \left(i\right)}\tilde{G}_{a}^{\left(i\right)}+s^{\left(i\right)}\tilde{S}^{\left(i\right)}+t_{1}^{\left(i\right)}\tilde{T}_{1}^{\left( i \right)}+t_{2}^{\left(i\right)}T_{2}^{\left(i\right)}+\bar{\psi}^{+\left(m\right)}\tilde{Q}^{+\left(m\right)}+\bar{\psi}^{-\left(m\right)}\tilde{Q}^{-\left(m\right)}+\bar{\rho}^{\left(m\right)}\tilde{R}
   ^{\left(m\right)}\,. \label{AGNH}
\end{equation}
Then, the three-dimensional CS action based on the super GNH$^{\left(N\right)}$ algebra is obtained by combining the gauge connection one-form \eqref{AGNH} and the non-vanishing components of the invariant tensor \eqref{invtGNH} into the general expression for a CS action \eqref{CS}, which thus yields
\begin{eqnarray}
 I_{\text{GNH}^{\left(N\right)}}&=& \alpha_{i}I_{i} \,\,=\, \alpha_0 I_0 + \alpha_1 I_1 + \ldots + \alpha_N I_N \,, \label{CSGNH}
\end{eqnarray}
where
\begin{eqnarray}
 I_{i}&=& \int \omega _{a}^{\left( j\right) }d\omega
^{a\left( k\right) }\delta _{j\ast k}^{i}+\epsilon ^{ac}\omega _{a}^{\left(
j\right) }\omega ^{\left( k\right) }\omega _{c}^{\left( l\right) }\delta
_{j\ast k\ast l}^{i}-2s^{\left( j\right) }d\omega ^{\left( k\right) }\delta
_{j\ast k}^{i}+2t_{1}^{\left(j\right)}dt_{2}^{\left(k\right)}\delta_{j\ast k}^{i}  \notag \\
&&+2\bar{\psi}^{-\left(m\right)}\nabla\psi^{-\left(n\right)}\delta_{m\ast (n-1)}^{i}+2\bar{\psi}^{+\left(m\right)}\nabla\rho^{\left(n\right)}\delta_{m\ast (n-1)}^{i}+2\bar{\rho}^{\left(m\right)}\nabla\psi^{+\left(n\right)}\delta_{m\ast (n-1)}^{i}\,. \label{IGNH}
\end{eqnarray}
In particular, the covariant derivatives of the spinor 1-forms for the GNH$^{\left(N\right)}$ superalgebra read
\begin{eqnarray}
 \nabla\psi^{+\left(m\right)}&=&d\psi^{+\left(m\right)}+\frac{1}{2}\omega^{\left(i\right)}\gamma_{0}\psi^{+\left(n\right)}\delta_{i\ast n}^{m}-\frac{1}{2}t_{1}^{\left(i\right)}\gamma_{0}\psi^{+\left(n\right)}\delta_{i\ast n}^{m} \,, \notag \\
 \nabla\psi^{-\left(m\right)}&=&d\psi^{-\left(m\right)}+\frac{1}{2}\omega^{\left(i\right)}\gamma_{0}\psi^{-\left(n\right)}\delta_{i\ast n}^{m}+\frac{1}{2}\omega^{a\left(i\right)}\gamma_{a}\psi^{+\left(n\right)}\delta_{i\ast n}^{m}+\frac{1}{2}t_{1}^{\left(i\right)}\gamma_{0}\psi^{-\left(n\right)}\delta_{i\ast n}^{m}\,, \notag \\
 \nabla\rho^{\left(m\right)}&=&d\rho^{\left(m\right)}+\frac{1}{2}\omega^{\left( i \right)}\gamma_{0}\rho^{\left(n\right)}\delta_{i \ast n}^{m}+\frac{1}{2}\omega^{a\left(i\right)}\gamma_{a}\psi^{-\left(n\right)}\delta_{i\ast n}^{m}+\frac{1}{2}s^{\left(i\right)}\gamma_{0}\psi^{+\left(n\right)}\delta_{i\ast n}^{m} \notag \\
 &&-\frac{1}{2}t_{2}^{\left(i\right)}\gamma_{0}\psi^{+\left(n\right)}\delta_{i\ast n}^{m}-\frac{1}{2}t_1^{\left(i\right)}\gamma_{0}\rho^{\left(n\right)}\delta_{i\ast n}^{m}\,. \label{GNHcurv}
\end{eqnarray}
The CS supergravity action \eqref{IGNH} is the most general CS action based on the GNH$^{\left(N\right)}$ superalgebra. In particular, as in the extended Newton-Hooke supergravity, the CS expression \eqref{CSGNH} can be split into two families: The CS term proportional to $\alpha_{2k}$ corresponds to NR exotic terms whose bosonic counterparts are related to the Pontryagin density; on the other hand, the contributions proportional to $\alpha_{2k+1}$ belong to the Euler CS family. In the particular case $N=2$, the CS supergravity action reproduces the EEB supergravity theory presented in \cite{Concha:2020tqx} by identifying the gauge field one-forms as
\begin{eqnarray}
 \omega^{\left(0\right)}&=&\omega\,,\qquad \qquad \omega_{a}^{\left(0\right)}=\omega_{a}\,,\qquad \quad \ \ \, s^{\left(0\right)}=s\,, \qquad \qquad  t_1^{\left(0\right)}=y_1\,,\qquad \qquad t_2^{\left(0\right)}=y_2\,, \notag \\
  \omega^{\left(1\right)}&=&\tau\,,\qquad \qquad \, \omega_{a}^{\left(1\right)}=e_{a}\,,\qquad \qquad s^{\left(1\right)}=m\,, \qquad \quad \ \, t_1^{\left(1\right)}=u_1\,, \qquad \qquad
    t_2^{\left(1\right)}=u_2\,,\notag \\
   \omega^{\left(2\right)}&=&k\,,\qquad \qquad \, \omega_{a}^{\left(2\right)}=k_{a}\,,\qquad \qquad s^{\left(2\right)}=t\,, \qquad \qquad t_1^{\left(2\right)}=b_1\,,\qquad \qquad \, t_2^{\left(2\right)}=b_2\,,\notag \\
   \psi^{+\left(1\right)}&=&\psi^{+}\,, \qquad \, \ \ \psi^{-\left(1\right)}=\psi^{-}\,, \qquad \quad \ \, \rho^{\left(1\right)}=\rho\,, \notag \\
   \psi^{+\left(2\right)}&=&\xi^{+}\,, \qquad \ \ \ \psi^{-\left(2\right)}=\xi^{-}\,, \qquad \quad \ \ \rho^{\left(2\right)}=\chi\,. 
\end{eqnarray}
The EEB supergravity action contains three independent sectors proportional to $\alpha_0$, $\alpha_1$, and $\alpha_2$. In such NR model, $I_0$ describes the NR exotic gravity coupled to the additional bosonic gauge fields $y_1$ and $y_2$. On the other hand, $I_1$ and $I_2$ describe a NR extended supergravity model in the presence of a cosmological constant and of extra bosonic content given by $k$, $k_{a}$, and $t$ (for further details about the EEB supergravity theory see \cite{Concha:2020tqx}). The $N=3$ case would now reproduce four independent sectors as in the GMEB supergravity theory previously introduced.

Let us note that, due to the non-degeneracy of the invariant tensor, the equations of motion of the GNH$^{\left(N\right)}$ supergravity theory are given by the vanishing of the associated curvature two-forms, which read
\begin{eqnarray}
F\left( \omega^{\left(i\right)} \right) &=&d\omega^{i}+\frac{1}{2}\bar{\psi}^{+\left(m\right)}\gamma^{0} \psi^{-\left(n\right)}\delta_{m\ast(n-1)}^{i} \,,  \notag \\
F^{a}\left( \omega ^{b\left(i\right)}\right) &=&d\omega^{a\left(i\right)}+\epsilon^{ac}\omega^{\left(j\right)}\omega_{c}^{\left(k\right)}\delta_{j\ast k}^{i}+\bar{\psi}^{+\left(m\right)}\gamma^{a} \psi^{-\left(n\right)}\delta_{m\ast(n-1)}^{i}\,, \notag\\
F\left(s^{\left(i\right)}\right)&=&ds^{\left(i\right)}+\frac{1}{2}\bar{\psi}^{-\left(m\right)}\gamma^{0} \psi^{-\left(n\right)}\delta_{m\ast(n-1)}^{i}+\bar{\psi}^{+\left(m\right)}\gamma^{0} \rho^{\left(n\right)}\delta_{m\ast(n-1)}^{i}\,, \notag\\
F\left(t_{1}^{\left(i\right)}\right)&=&dt_{1}^{\left(i\right)}+\frac{1}{2}\bar{\psi}^{+\left(m\right)}\gamma^{0}\psi^{+\left(n\right)}\delta_{m\ast(n-1)}^{i}\,, \notag \\
F\left(t_{2}^{\left(i\right)}\right)&=&dt_{2}^{\left(i\right)}-\frac{1}{2}\bar{\psi}^{-\left(m\right)}\gamma^{0} \psi^{-\left(n\right)}\delta_{m\ast(n-1)}^{i}+\bar{\psi}^{+\left(m\right)}\gamma^{0} \rho^{\left(n\right)}\delta_{m\ast(n-1)}^{i}\,, \label{curvignh}
\end{eqnarray}
along with \eqref{GNHcurv}.

It is interesting to notice that the GNH$^{(N)}$ superalgebras are related to the GEB$^{\left(N\right)}$ ones through an IW contraction procedure. The following diagram summarizes the expansion and contraction relations:
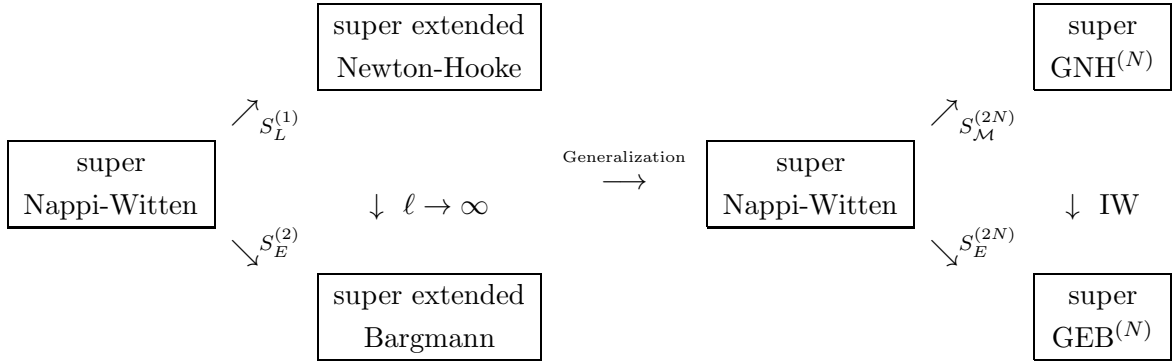
\begin{equation*}
\begin{tabular}{ccc}
\cline{3-3}
&  & \multicolumn{1}{|c|}{super extended} \\
&  & \multicolumn{1}{|c|}{Newton-Hooke} \\ \cline{3-3}
& $\nearrow _{S_{L}^{\left( 1\right) }}$ &  \\ \cline{1-1}
\multicolumn{1}{|c|}{super} \\
\multicolumn{1}{|c}{Nappi-Witten}& \multicolumn{1}{|c}{} & $\downarrow $ \
$\ell \rightarrow \infty $ \\ \cline{1-1}
& $\searrow ^{S_{E}^{\left( 2\right) }}$ &  \\ \cline{3-3}
&  & \multicolumn{1}{|c|}{super extended} \\
&  & \multicolumn{1}{|c|}{Bargmann} \\ \cline{3-3}
\end{tabular}%
\overset{%
\begin{array}{c}
\text{{\tiny Generalization}}%
\end{array}%
}{\longrightarrow }%
\begin{tabular}{ccc}
\cline{3-3}
&  & \multicolumn{1}{|c|}{super}\\
&  & \multicolumn{1}{|c|}{GNH$^{\left(N\right)}$} \\
 \cline{3-3}
& $\nearrow _{S_{\mathcal{M}}^{\left( 2N\right) }}$ &  \\ \cline{1-1}
\multicolumn{1}{|c|}{super} \\
\multicolumn{1}{|c}{Nappi-Witten}& \multicolumn{1}{|c}{} & $\downarrow$ \ IW \\ \cline{1-1}
& $\searrow ^{S_{E}^{\left( 2N\right) }}$ &  \\ \cline{3-3}
&  & \multicolumn{1}{|c|}{super}\\
&  & \multicolumn{1}{|c|}{GEB$^{\left(N\right)}$} \\
\cline{3-3}
\end{tabular}%
\end{equation*}
In particular, the GEB$^{\left(N\right)}$ superalgebras appear as an IW contraction of the generalized extended Newton-Hooke ones by rescaling the super GNH$^{\left(N\right)}$ generators as follows:
\begin{eqnarray}
 \tilde{J}^{\left(i\right)}&\rightarrow&\sigma^{2i}\tilde{J}^{\left(i\right)}\,, \qquad \qquad \qquad \tilde{Q}_{\alpha}^{+\left(m\right)}\rightarrow\sigma^{2m-1}\tilde{Q}_{\alpha}^{+\left(m\right)}\,,\notag \\
 \tilde{G}_{a}^{\left(i\right)}&\rightarrow&\sigma^{2i}\tilde{G}_{a}^{\left(i\right)}\,,\qquad \qquad \quad \ \ \ \tilde{Q}_{\alpha}^{-\left(m\right)}\rightarrow\sigma^{2m-1}\tilde{Q}_{\alpha}^{-\left(m\right)}\,, \notag \\ 
 \tilde{S}^{\left(i\right)}&\rightarrow&\sigma^{2i}\tilde{S}^{\left(i\right)}\,, \qquad \qquad \qquad \ \ \, \tilde{R}_{\alpha}^{\left(m\right)}\rightarrow\sigma^{2m-1}\tilde{R}_{\alpha}^{\left(m\right)}\,,
\end{eqnarray}
and considering the limit $\sigma\rightarrow\infty$. Note that the vanishing cosmological constant limit $\ell\rightarrow\infty$ performed in the $N=1$ case to recover the extended Bargmann superalgebra can also be seen as an IW contraction. At the CS action level, the proper IW contraction requires rescaling, in addition, the NR parameters in the invariant tensor \eqref{invtGNH} as $\alpha_{i}\rightarrow\sigma^{i}\alpha_{i}$.

As an ending remark, let us note that the $S_{E}$ and $S_{\mathcal{M}}$ semigroups used to obtain the respective generalizations of the extended Bargmann and extended Newton-Hooke supergravity theories are the same used to find generalizations of the super Poincaré and super AdS algebra. Furthermore, as it was shown in \cite{Concha:2020sjt,Penafiel:2019czp}, the same semigroups were considered at the bosonic level, allowing to define the GEB$^{\left(N\right)}$ and GNH$^{\left(N\right)}$ algebras.


\section{Concluding remarks}\label{concl}

In this paper, we have introduced an alternative procedure to obtain diverse NR CS supergravity theories in three spacetime dimensions. Known and new NR superalgebras have been obtained considering the expansion method based on semigroups, the so-called $S$-expansion, to a Nappi-Witten superalgebra introduced in \cite{Concha:2020tqx}. Interestingly, the $S$-expansion allows to immediately obtain the non-vanishing components of the invariant tensor of an expanded superalgebra in term of the original ones. Such advantage has allowed us to construct, in a systematic way, the respective NR CS supergravity actions for each NR superalgebra presented. 

We have shown that our resulting theories can be split into two NR supergravity families. Indeed, the extended Bargmann supergravity along with its Maxwellian version can be seen as particular subcases of a generalized extended Bargmann supergravity theory which we have denoted as GEB$^{\left(N\right)}$ supergravity. In particular, for $N=1$ we recover the extended Bargmann supergravity theory. On the other hand, the extended Newton-Hooke supergravity belongs to a generalized extended Newton-Hooke theory which we have called as GNH$^{\left(N\right)}$. Such generalizations correspond to supersymmetric extensions of the GEB$^{\left(N\right)}$ and GNH$^{\left(N\right)}$ bosonic algebras recently introduced in \cite{Concha:2020sjt}. Remarkably, both families are related through an IW contraction process, similarly as their bosonic counterparts.

It would be worth considering further studies on the Maxwellian version and generalizations of the extended Bargmann supergravity theory. One could analyze, for instance, the Schrödinger extension \cite{Afshar:2015aku,Bergshoeff:2015ija} of the NR generalized superalgebras presented here in a similar way as it was done in the case of the extended Schrödinger supergravity \cite{Ozdemir:2019tby}. In particular, the map between Newton-Cartan geometry and Horava-Lifshitz gravity \cite{Hartong:2016yrf} could suggest a superconformal non-projectable Horava-Lifshitz gravity. The Schrödinger version of our results would allow us to approach an off-shell formulation of the respective NR supergravity actions which could serve to construct NR effective field theories on curved backgrounds by means of localization \cite{Pestun:2007rz,Festuccia:2011ws}. On the other hand, it would be intriguing to explore the possibility to apply the $S$-expansion method in the context of the Schrödinger superalgebra families, in order to establish a systematic way to obtain generalized Schrödinger supergravity actions, in a very similar way to the construction presented here. 

A future development could also consist in generalizing our results to the extended Newtonian family. One could expect to obtain novel Newtonian supergravity theories, different from the extended Newtonian one presented in \cite{Ozdemir:2019orp}, being supersymmetric extensions of the exotic and Maxwellian extended Newtonian gravity recently introduced in \cite{Concha:2019dqs} and \cite{Concha:2020ebl}, respectively. Newtonian gravity models are worth studying as they offer an action principle for Newtonian gravity through the CS formalism different from the one introduced in \cite{Hansen:2018ofj}. It would be interesting to study the matter coupling of the new Newtonian supergravity theories as well.

The procedure considered here could be extended to the ultra-relativistic (UR) regime. In particular, the construction of UR supergravity models remains poorly explored \cite{Ravera:2019ize,Ali:2019jjp}. The $S$-expansion method could be applied to a Carrollian version of the Nappi-Witten symmetry to obtain new UR superalgebras (work in progress). One could expect to find two UR superalgebra families being the respective UR versions of the relativistic $\mathfrak{B}_{N+2}$ and $\mathfrak{C}_{N+2}$ superalgebras. At the bosonic level, the Carrollian symmetries emerge in the framework of flat holography and fluid/gravity correspondence \cite{Ciambelli:2018xat,Ciambelli:2018wre,Ciambelli:2018ojf,Campoleoni:2018ltl}, whose applications motivate us to explore supersymmetric extensions of the Carrollian symmetries in the context of supergravity.


\section*{Acknowledgment}

This work was funded by the National Agency for Research and Development ANID (ex-CONICYT) - PAI grant No. 77190078 (P.C.) and the FONDECYT Project N$^{\circ}$3180594 (M.I.) and N$^{\circ }$3170438 (E.R.). P.C. would like to
thank to the Dirección de Investigación and Vice-rectoría de Investigación
of the Universidad Católica de la Santísima Concepción, Chile, for their
constant support. L.R. would like to thank the Department of Applied Science and Technology of the Polytechnic University of Turin, and in particular F. Dolcini and A. Gamba, for financial support.


\bibliographystyle{fullsort}
 
\bibliography{NR_supergravity_and_semigroup_expansion_vf}

\end{document}